\documentclass{article}

    \PassOptionsToPackage{numbers, compress}{natbib}
    \usepackage[final]{neurips_2024}

\usepackage[utf8]{inputenc} % allow utf-8 input
\usepackage[T1]{fontenc}    % use 8-bit T1 fonts
\usepackage{hyperref}       % hyperlinks
\usepackage{url}            % simple URL typesetting
\usepackage{booktabs}       % professional-quality tables
\usepackage{amsfonts}       % blackboard math symbols
\usepackage{nicefrac}       % compact symbols for 1/2, etc.
\usepackage{microtype}      % microtypography
\usepackage{xcolor}         % colors

\usepackage{minitoc}
\usepackage{wrapfig}
\usepackage{makecell}
\usepackage{diagbox}
\usepackage{graphicx}
\usepackage{verbatim}
\usepackage{multirow}
\usepackage{algorithmic}
\usepackage{hyperref}
\usepackage{longtable}
\usepackage{array}
\usepackage{tocloft}
\makeatletter
\let\c@lofdepth\relax
\let\c@lotdepth\relax
\makeatother
\usepackage{subfigure}
\usepackage{stfloats}
\usepackage{multicol}
\usepackage{color}
\usepackage{bm}
\usepackage{amsmath}
\usepackage{booktabs} % For formal tables
\usepackage{epsfig}
\usepackage{enumitem}
\usepackage{cleveref}
\usepackage{arydshln}
\usepackage{balance}
\usepackage{xspace}
\usepackage{enumitem}
\usepackage{listings}
\usepackage{xcolor}
\usepackage{appendix}
\usepackage{imakeidx}
\makeindex[name=app, title=Appendix Index]

\definecolor{codegreen}{rgb}{0,0.6,0}
\definecolor{codegray}{rgb}{0.5,0.5,0.5}
\definecolor{codepurple}{rgb}{0.58,0,0.82}
\definecolor{backcolour}{rgb}{1,1,1}

\usepackage{xcolor}
\definecolor{blue-violet}{rgb}{0.54, 0.17, 0.89}
\hypersetup{colorlinks=true, linkcolor=blue-violet, citecolor=blue, urlcolor=blue-violet}
\newcommand{\tabincell}[2]{\begin{tabular}{@{}#1@{}}#2\end{tabular}}
\newcommand{\vpara}[1]{\vspace{0.05in}\noindent \textbf{#1 }}
\newcommand{\ipara}[1]{\vspace{0.03in}\noindent \textit{#1 }}

\newcommand{\beq}[1]{\vspace{-0.03in}\begin{equation}#1\end{equation}\vspace{-0.03in}}
\newcommand{\beqn}[1]{\vspace{-0.04in}\begin{eqnarray}#1\end{eqnarray}\vspace{-0.04in}}

\newtheorem{problem}{Problem}
\newtheorem{definition}{Definition}

\newcommand{\model}{MSAGPT}
\newcommand{\smodel}{MSAGPT\space}
\title{MSAGPT: Neural Prompting Protein Structure Prediction via MSA Generative Pre-Training}

\author{
Bo Chen$^{1\ast\dagger}$, Zhilei Bei$^{1\ast\dagger}$, Xingyi Cheng$^{2}$, Pan Li$^{2}$, Jie Tang$^{1}$,  Le Song$^{2,3}$\\
 $^1$\textmd{Tsinghua University}
$^2$\textmd{BioMap Research}
$^3$\textmd{MBZUAI}\\
\texttt{cb21@mails.tsinghua.edu.cn} \\
 \normalsize\rule{0pt}{1em}\url{https://github.com/THUDM/MSAGPT}\\
}

\begin{document}

\doparttoc
\faketableofcontents
\maketitle

\renewcommand{\thefootnote}{\fnsymbol{footnote}}
    \footnotetext[1]{BC and ZB contributed equally.}
    \footnotetext[2] {Work done while interned at BioMap.}

\begin{abstract}

Multiple Sequence Alignment (MSA) plays a pivotal role in unveiling the evolutionary trajectories of protein families. 
% The accuracy of protein structure predictions is often compromised in these protein sequences lacking sufficient homologous information to construct high-quality MSA. 
The accuracy of protein structure predictions is often compromised for protein sequences that lack sufficient homologous information to construct high-quality MSA. 
Although various methods have been proposed to generate virtual MSA under these conditions, they fall short in comprehensively capturing the intricate co-evolutionary patterns within MSA or require guidance from external oracle models. 
Here we introduce \model, a novel approach to prompt protein structure predictions via MSA generative pre-training in the low-MSA regime. \smodel employs a simple yet effective 2D evolutionary positional encoding scheme to model the complex evolutionary patterns. Endowed by this, its flexible 1D MSA decoding framework facilitates zero- or few-shot learning.
% combined with a 1D zero-shot or few-shot MSA decoding framework, enabling comprehensive modeling of co-evolutionary patterns and facilitating flexible MSA decoding. 
Moreover, we demonstrate that leveraging the feedback from AlphaFold2 can further enhance the model's capacity via Rejective Fine-tuning (RFT) and Reinforcement Learning from AF2 Feedback (RLAF). 
Extensive experiments confirm the efficacy of \smodel in generating faithful virtual MSA to enhance the structure prediction accuracy (up to +8.5\% TM-Score on few-shot scenarios). The transfer learning capabilities also highlight its great potential for facilitating other protein tasks.
% relying on the quality of MSA.
% The corresponding code, data, and script will be available at.
% advance the field of protein structure prediction.
\end{abstract}
\section{Introduction}
\label{sec:intro}
% The advancement of deep learning techniques has shown tremendous success in tackling many critical yet long-standing science problems, such as AlphaFold2~\cite{}, for protein tertiary structure prediction, AlphaGeometry~\cite{} for geometry math reasoning, PanGu~\cite{} for weather forecasting, etc. Especially for the area of structural biology, the rise of Alphafold2 marks the significant milestone where the in-vitro accuracy of protein structure prediction have matched that of the wet-lab experiments (~90\% atomic accuracy). The success of Alphafold2 largely lies in the usage of co-evolutionary information supported by the Multiple Sequence Alignments (MSA). 
% MSA, derived from querying a protein sequence
% against vast databases using search algorithms, represent aggregations of homologous sequences
% that capture evolutionary information, acting as the foundation of many PSP models. 
% However, not all protein sequences possess a rich set of homologous counterparts. This scarcity often means that even advanced search algorithms struggle to construct high-quality MSA, leading to a compromised efficacy for MSA-reliant models like AF2.
% which exemplifies the profound impact of these technologies.
% achieving an in-vitro precision for predicting protein structures that rivals traditional wet lab experiments, with approximately 90\% atomic accuracy. 

\begin{figure}[t]
	\centering
	\subfigure[A toy example of MSA.]{\label{subfig:msa}
		\includegraphics[width=0.45\textwidth]{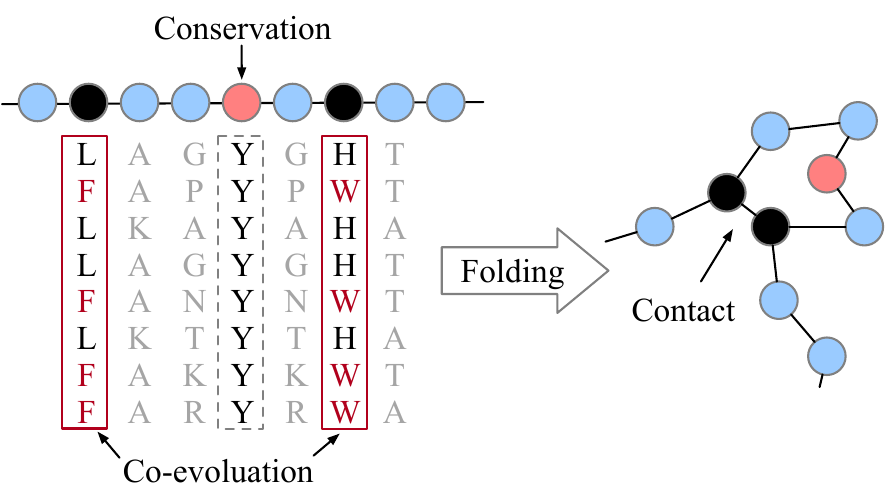}
	}
	\hspace{-0.1in}
	\subfigure[Overall performance comparisons.]{\label{subfig:per}
		\includegraphics[width=0.40\textwidth]{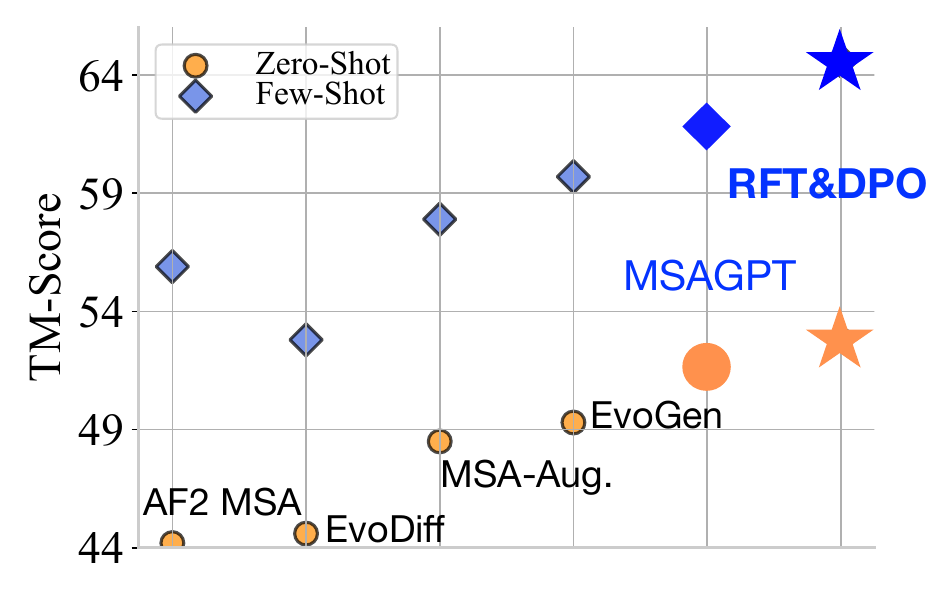}
	}
	% \vspace{-10pt}
	\caption{\label{fig:motiv} \textbf{(a) The illustration of MSA and (b) performance comparisons between \smodel and advanced baselines on three natural MSA-scarce benchmark.}}
\end{figure}

The advent of deep learning has significantly propelled progress across various scientific domains, exemplified by breakthroughs such as AlphaFold series~\cite{abramson2024accurate, jumper2021highly} for accurate biomolecular interaction predictions, AlphaGeometry~\cite{trinh2024solving} for intricate geometry and mathematical reasoning——to name a few.
Among these, AlphaFold2 (AF2) represents a landmark within structural biology, achieving an \textit{in silico} precision of approximately 90\% atomic accuracy that rivals wet lab experiments on protein structure predictions (PSP). The remarkable success of AF2 can be attributed to its innovative end-to-end use of co-evolutionary information supported by Multiple Sequence Alignment (MSA).
MSA aggregates homologous sequences from vast databases, providing a comprehensive overview of evolutionary trajectories, which is critical for accurately predicting protein structures~\cite{abramson2024accurate, jumper2021highly, baek2021accurate}. 
An illustrative example in Figure~\ref{fig:motiv}(a) showcases that the correlations analysis among amino acids (AAs) sites could reveal contacts or conservative regions in the folding structure. 
% However, the landscape of protein sequences is marked by variability in the availability of homologous counterparts. 
Unfortunately, not all proteins possess a rich set of homologous counterparts.
Statistical investigations reveal that approximately 20\% of metagenomic proteins~\cite{pearson2013introduction} and around 11\% of proteins from eukaryotic and viral origins~\cite{perdigao2015unexpected} are classified as "orphan" proteins. 
% which inherently lack sequence homologs. 
% This scarcity of homologs presents a significant challenge for advanced search algorithms in constructing high-quality MSA, consequently impeding the performance of MSA-based PSP models. 
This presents a significant challenge for MSA-search algorithms in constructing high-quality MSA, consequently impeding the performance of PSP models~\cite{jumper2021highly}. 
% A toy example is shown in Figure~\ref{fig:motiv}(a), the pairwise correlations among amino acid sites can deduce contacts within protein folding structures.

% Nevertheless, not all protein sequences possess a rich set of homologous counterparts. Statistically, about 1/5 of all metagenomic protein
% sequences~\cite{pearson2013introduction} and about 11\% of eukaryotic and viral proteins~\cite{perdigao2015unexpected} are estimated to be ``orphan'' which naturally lack sequence homologs. This scarcity impedes the ability of advanced search algorithms to construct high-quality MSA, directly impacting the efficacy of MSA-based PSP models. 

% The scarcity of homologous sequences for these orphan proteins underscores a critical bottleneck in leveraging evolutionary information for accurate structure prediction, highlighting a key area where current methodologies may fall short.

% \begin{figure}[t]
%     \centering
%     \includegraphics[width=0.5\textwidth]{figures/motivation.pdf}
%     \caption{\textbf{illustrate the MSA and show the final comparison results.}}
%     \label{fig:motiv}
% \end{figure}

Drawing on the impressive capabilities of large language models endowed either by the autoencoding~\cite{devlin2018bert} or the autoregressive language modeling regime~\cite{bubeck2023sparks, touvron2023llama}, protein language models (PLMs) have been developed to unveil the evolutionary patterns and sequence characteristics intrinsic to protein structures.
Specifically, generative PLMs~\cite{nijkamp2023progen2, chen2024xtrimopglm, ferruz2022protgpt2}, trained on vast protein databases~\cite{suzek2007uniref, berman2000protein, steinegger2019protein, steinegger2018clustering} 
% containing billions of sequences, 
have achieved unparalleled success in generating novel proteins with desired structural properties. These achievements underscore the efficacy of language models in identifying evolutionary patterns within individual protein sequences.
Inspired by this, subsequent works~\cite{rao2021msa, zhang2021co} attempt to further integrate MSA as the input 
% with single protein sequence inputs 
or by directly generating virtual yet informative MSA~\cite{truong2024poet, zhang2023enhancing, zhang2023unsupervisedly}
% \cxy{Figure 1(a) seems that have little relation with this motivation. It's a little bit confusing.}
to provide additional evolutionary insights.
% thereby improving the accuracy of PSP models. 
These approaches usually adopt customized attentions that merely allow attention aggregated among specific directions, such as axial attention~\cite{ho2019axial}, for separately analyzing the row- and column-wise co-evolutionary patterns in MSA.  
% As Figure~\ref{fig:motiv}(a) illustrated that one must analyse the pairwise relationships of amino acid sites across all homogeneous sequences concurrently that conclude the structural constraint on the folding structures.
However, these attention mechanisms usually have low efficiency in capturing the evolutionary information in MSA, or even fail to adequately capture intricate co-evolutionary dynamics.
Taking Figure~\ref{fig:motiv}(a) as an example, it is imperative to concurrently analyze the pairwise or high-order relationships of amino acid sites across all homologs to deduce the structural constraints influencing the folding structures, which may not achieved by customized attention. 
% \vspace{-10pt}
The limited capacity may result in compromised performance on the task that highly resorts to co-evolutionary information.
% \vspace{-10pt}
% a novel Transformer-based architecture termed with the Evolutionary-Positional-Encoding mechanism, 
% \cxy{pioneering?, which prompts ...}
% \vpara{Present Work.} 
Built upon the insights mentioned above, 
we introduce \model, 
a simple yet effective framework that prompts protein structure prediction via MSA generative pre-training.
This method facilitates \textit{de novo} MSA generation, aiding in protein structure prediction in scenarios with limited MSA available. \smodel is characterized by its unique features:
% to perform \textit{de novo} MSA generation for prompting the protein tertiary structure prediction in contexts where limited co-evolutionary information is available.
% Specifically, \smodel has the following characteristics:

\noindent $\bullet$ \textbf{2D Evolutionary Positional Encoding.} 
We employ an innovative dual-axis positional encoding scheme that captures column- and row-wise co-evolutionary information concurrently. This method provides a comprehensive understanding of complex evolutionary relationships with high efficacy. 
enhancing the model's generative capabilities.
% We have meticulously designed a dual-axis positional encoding scheme that considering the column- and row-wise co-evolutionary information simultaneously. 
% This positional encode mechanism enables a more comprehensive characterization of evolutionary relationships.

\noindent $\bullet$ \textbf{1D Zero-/Few-Shot MSA Decoding.} 
With 2D positional encoding, \smodel re-formalizes MSA generation as a one-dimensional sequence generation task, optimized by the simple next-token-prediction objective. This enables \smodel to conduct zero- or few-shot MSA generation under a flexible in-context learning framework.
% and compatible with high-performance computational techniques used in large-scale language models.
% Leveraging the 2D positional encoding framework, \smodel reformulates the MSA generation task into a one-dimensional sequence generation format, which is optimized by simple next-token-predictions pre-training objectives.  This allow \smodel is not only capable of performing zero-shot or few-shot MSA generation within a flexible in-context learning paradigm, but also is fitted with any high-performance computation framework employed in large language models.

\noindent $\bullet$ \textbf{Learning from AlphaFold2 Feedback.} 
\smodel further utilizes feedback from AlphaFold2 to reduce hallucinations during MSA generation. 
This approach ensures the generation of reliable and informative MSA, thus enhancing protein structure prediction.

% \smodel further learns from AlphaFold2 feedback. as the gold standard severed as the reward model to guide the model to reduce hallucinations during MSA generation. Such that it can generate faithful and informative MSA to enhance the performance of protein structure predictions.

Extensive experiments conducted on three benchmarks, CAMEO~\cite{haas2018continuous}, CASP, and PDB~\cite{berman2000protein}, demonstrate the superior capacity of \smodel in generating high-quality MSA.
Notably, \smodel outperforms existing MSA generation models on both zero- and few-shot scenarios.
Impressively, AF2 with virtual MSA generated by \smodel significantly improves the structure prediction accuracy than that with natural MSA on cases with limited homologous information. 
Moreover, the subsequent Rejective Fine-tuning (RFT) and learning from AF2 feedback (RLAF) stage further enhance the model's ability to generate informative and faithful MSA, outperforming the original \smodel by a large margin, as shown in Figure~\ref{subfig:per}.
Additionally, we demonstrate that virtual MSA can also benefit other tasks.
We expect \smodel to become integral in supplementing protein-related tasks requiring critical evolutionary information from MSA. The model is available at \href{https://github.com/THUDM/MSAGPT}{https://github.com/THUDM/MSAGPT}.
\section{Related work}
\label{sec:rela_work}
% Here, we recall the prevailing MSA generation methods and the post-alignment techniques with reinforcement learning.

\begin{figure*}[t]
    \centering
    \includegraphics[width=\textwidth]{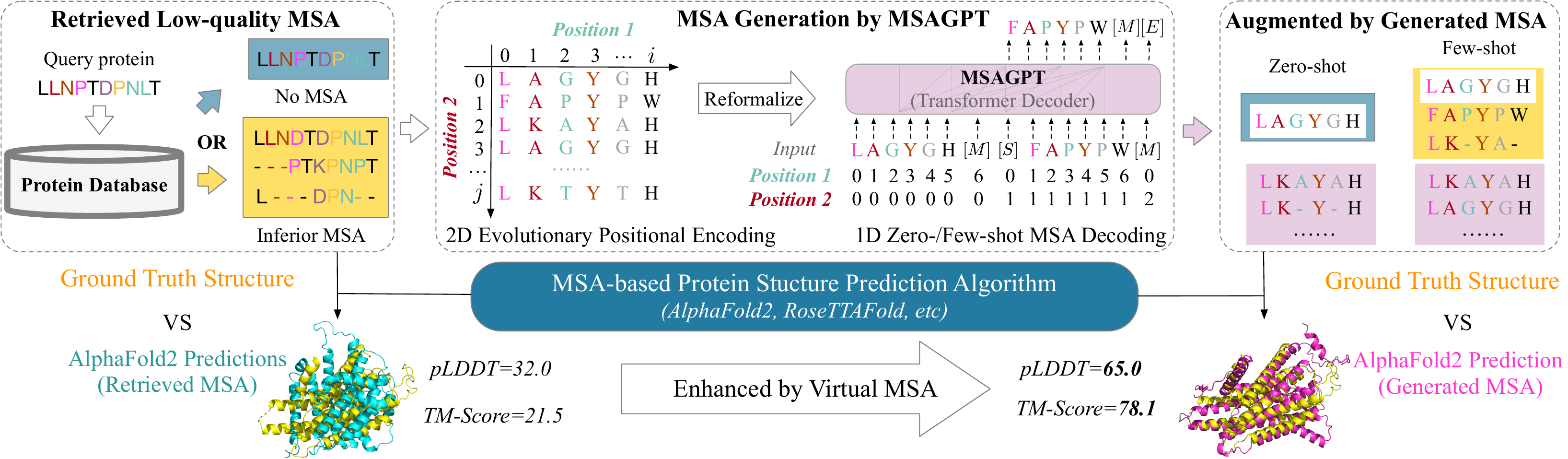}
    \caption{\label{fig:overall_frame} \textbf{The overall framework of prompting protein structure predictions via MSA generation.}
\textmd{\textbf{Left}: The challenge faced by conventional search algorithms on protein with scarce homologous sequences, resulting in suboptimal alignments. 
\textbf{Middle-to-Right}: \smodel generates informative and high-quality MSA for such challenging queries, presenting a promising approach to overcoming these limitations.
$\texttt{[M]}$ denotes the sequence separator. $\texttt{[S]}, \texttt{[E]}$ are the special tokens to represent the start or end of MSA generation. 
}
    }
\end{figure*}

\vpara{Protein Structure Prediction.}
Proteins are fundamental to the various biological processes that sustain, grow, and protect living organisms. 
% Deciphering the biological information embedded in proteins is essential for understanding the complex mechanisms of life. 
% Given that protein sequences dictate their structure and function~\cite{anfinsen1959molecular},
Groundbreaking deep learning approaches~\cite{abramson2024accurate, jumper2021highly, baek2021accurate} have been developed to predict the folding structures based on their sequences. These methods have achieved structure prediction accuracy to conventional wet-lab experiments. The success largely relies on the utilization of MSA, which are retrieved through search algorithms~\cite{zhang2020deepmsa, zheng2024improving, johnson2010hidden, steinegger2017mmseqs2} against vast databases~\cite{suzek2007uniref, berman2000protein, steinegger2019protein, steinegger2018clustering}. However, challenges arise with ``orphan'' protein sequences, which lack sufficient homologous sequences for accurate structure prediction. 
Single-sequence PSP methods~\cite{chen2024xtrimopglm, lin2023evolutionary, wu2022high, chowdhury2022single} are designed to infer folding structures directly from the query protein sequences. Despite these advancements, the accuracy of predictions from single-sequence methodologies generally falls short of those derived from MSA-based algorithms.

% In response to these challenges, our research aims to bridge this gap by generating virtual yet effective MSA to enhance protein structure prediction, particularly in scenarios where available MSA are of inferior quality. This innovative approach seeks to leverage the strengths of both single-sequence and MSA-based methodologies, potentially offering a new paradigm in the accurate prediction of protein tertiary structures under constrained conditions.

\vpara{Protein Language Models.}
Protein Language Models (PLMs), such as ESM~\cite{lin2023evolutionary, rives2021biological}, ProGen~\cite{nijkamp2023progen2, madani2023large}, etc~\cite{ferruz2022protgpt2, elnaggar2021prottrans, alamdari2023protein} have emerged as a groundbreaking development in computational biology. PLMs are trained on single sequences, towards understanding protein structural features or enabling the generation of diverse and realistic protein sequences. MSA Transformer~\cite{rao2021msa} further incorporates MSA as the input, achieving better performance than these single sequence models, underscoring the importance of utilizing the evolutionary information from MSA~\cite{frazer2021disease, chen2023improved, sturmfels2020profile}.
% on par with the ESM2-15B model in protein structure prediction tasks with merely 100M parameters. This highlights the critical advantage of integrating co-evolutionary information from MSA, 
% As for protein design, PLMs including ProGen~\cite{ nijkamp2023progen2, madani2023large} and ProtGPT2~\cite{ferruz2022protgpt2}, endowed by the autoregressive training regime, enable the generation of diverse and realistic protein sequences.
To enhance MSA generation, MSA-Augmentor~\cite{zhang2023enhancing}, PoET~\cite{truong2024poet} employ the seqs2seqs pre-training, 
% framework with both encoder and decoder as the MSA Transformer. 
% This allows for the augmentation of MSA using inferior searched MSA as few-shot prompts. 
which adopts the sequential axial attention mechanism to capture the evolutionary data across and within the input sequences, both horizontally and vertically.
EvoGen~\cite{zhang2023unsupervisedly}, serving as the meta generative model, aims at producing virtual MSA for enhancing protein structure predictions. However, it highly resorts to external structural prediction models to optimize its performance.
\vpara{Aligning with Human Preferences.} Aligning language models with human preferences has been shown to be effective in improving generation quality~\cite{bubeck2023sparks, rafailov2023direct, schulman2017proximal, team2023gemini}. 
% Learning from human feedback based on pre-trained models is a common approach in achieving this alignment.
Existing methods typically employ supervised fine-tuning using human-annotated datasets or reinforcement learning from human feedback pipelines~\cite{rafailov2023direct, schulman2017proximal}. 
% Reinforcement algorithms such as Proximal Preference Optimization~\cite{schulman2017proximal} (PPO) and Direct Preference Optimization~\cite{rafailov2023direct} (DPO) are commonly used in these pipelines.
Inspired by these, we utilize the feedback from AlphaFold2 to further enhance the generation capability of the pre-trained model, which helps mitigate hallucinations and enables the model to generate accurate and reliable MSA.
% judge whether the generated virtual MSA is high-quality or not. Guided by the AF2-preference signals, our model tends to generate the MSA containing clean and sufficient MSA rather 
% than decaying to hallucination circumstance like merely replicating segments of the input query sequence.
% Inspired by this, we treat AlphaFold2 as the reward model to further guide the pre-trained model for alleviating generating hallucination and generating faithful MSA.
\section{Preliminary}
\label{sec:prob}
% Here we formalize the esseb definition of multiple sequence alignment (MSA) and the problem of MSA generation.
\begin{definition}
	\textbf{Multiple Sequence Alignment (MSA)}. Given the query protein sequence $Q \in \mathbb{A}^{1 \times L}$, where $\mathbb{A}$ denotes the set of alphabetic symbols used to represent the 20 basic amino acids and $L$ represents the number of amino acids per sequence, the MSA $M \in \mathbb{A}^{N \times L}$ of $Q$ is comprised of $N$ homologous protein sequences, which can be obtained either by searching over protein databases or generating with MSA generation methods. 
\end{definition}

% Noted that the MSA $M$ can be obtained either by retrieval over protein database or generated by MSA generation methods.

\begin{problem}
	\textbf{Prompting Protein Structure Prediction by MSA Generation}.  Given $Q$ with initial MSA $M_{\text{init}} \in \mathbb{A}^{n \times L}$ as the prompt, where $n=0$ indicates the zero-shot MSA generation and $n > 0$ signifies the few-shot MSA generation, we target at learning a function $f$ to generate virtual MSA $M_{\text{gen}} \in \mathbb{A}^{m \times L}$ based on $Q$ and $M_{\text{init}}$, such that the structure prediction accuracy based on the augmented MSA $M_{\text{aug}} \in \mathbb{A}^{(n+m) \times L}$ significantly surpasses that based on the initial MSA $M_{\text{init}}$,
	\beqn{
		M_{\text{aug}} \!\!\!\!&=&\!\!\!\! f(Q, M_{\text{init}}), \nonumber \\ 
        \mathbb{I}_{acc}(Q, M_{\text{aug}}) \!\!\!\!&>&\!\!\!\! \mathbb{I}_{acc}(Q, M_{\text{init}})  \nonumber
	}

\noindent where the $\mathbb{I}_{acc}$ is prediction accuracy comparing the prediction result of AF2 and the ground truth. 
% \noindent where the $\mathbb{I}_{acc}$ is prediction accuracy determined by comparing the prediction result of AF2 and the ground truth. 
% Unless otherwise mentioned, we use TM-Score~\cite{} as the metric in this paper, which is widely used in the structure biology domain to measure the similarity between two three dimensional protein stucture. 
\end{problem}

In this paper, we mainly focus on improving the structure prediction accuracy in the low-MSA regime, i.e., the cases that lack a sufficient number of homologous sequences.

% In this paper, we mainly focus on improving the structure prediction accuracy in cases with limited co-evolutionary information where the MSA obtained from protein databases lack a sufficient number of homologous sequences.
% In this paper, our primary focus on enhancing the accuracy of protein structure prediction in cases characterized by limited co-evolutionary information. Specifically, we address scenarios where the Multiple Sequence Alignments (MSAs) obtained from protein databases lack a sufficient number of homologous sequences.
% Surprisingly, we also observe that the virtual MSA generated by our model also boost the cases with sufficient MSA number. Even merely with our generated MSA, AF2 can accurately predicts the structure similarly with that with searched MSA. 

% \begin{figure*}[t]
%     \centering
%     \includegraphics[width=\textwidth]{figures/overall_frame}
%     \caption{\label{fig:overall_frame} \textbf{The overall framework of prompting protein structure predictions via MSA generation.}
% \textmd{\textbf{Left}: The challenge faced by conventional search algorithms on protein with scarce homologous sequences, resulting in suboptimal alignments. 
% \textbf{Middle-to-Right}: \smodel generates informative and high-quality MSA for such challenging queries, presenting a promising approach to overcoming these limitations.
% $\texttt{[M]}$ denotes the sequence separator. $\texttt{[S]}, \texttt{[E]}$ are the special tokens to represent the start or end of MSA generation. 
% }
%     }
% \end{figure*}

\section{Methodology}
\label{sec:method}
% \todo{1. Add discussion about the training or inference efficiency about \model.}
Given a query sequence and its retrieved natural MSA, we aim to comprehensively understand the co-evolutionary patterns in MSA,
% imposed by the pairwise inter-residue correlations among MSA. 
such that we can generate informative virtual MSA for prompting protein structure prediction in the low-MSA regime. 
Conceptually, the co-evolutionary information is analogous to the covariance matrix in mathematics, depicting the correlations among amino acids by comparing pairwise or high-order correlations among amino acid sites in MSA, as depicted in Figure~\ref{fig:motiv}(a).
% Thus, our model should capture the pairwise relationships among rows and columns simultaneously to offer a comprehensive insight into evolutionary information. 
To achieve this goal, \smodel contains two key adoptions, distinguishing it from existing MSA-based PLMs that rely on customized attentions~\cite{jumper2021highly, rao2021msa, zhang2023enhancing, truong2024poet}: 
\textbf{2D Evolutionary Positional Encoding.} Introduces an adaptive dual-axis positional encoding scheme that captures column- and row-wise co-evolutionary information concurrently. 
And
\textbf{1D Zero-/Few-Shot MSA Decoding.} Re-formalizes MSA generation as a one-dimensional sequence generation task based on the proposed 2D positional encoding scheme, which enables \smodel to conduct zero- or few-shot context learning MSA generation framework.
The overall framework is illustrated in Figure~\ref{fig:overall_frame}.

% \begin{figure}[t]
%     \centering
%     \includegraphics[width=0.5\textwidth]{figures/framework2.pdf}
%     \caption{\textbf{illustrate the MSA and show the final comparison results.}}
%     \label{fig:task}
% \end{figure}
% $x_{i,j}$ represents the residue located at the $i$-th row and $j$-th column in $\mathbf{X}$. 

% \cxy{We need to explain why 2D positional encoding is better than axis-attentions from biological perspectives.}
\subsection{2D Evolutionary Positional Encoding}
Vanilla transformers typically use 1D positional embeddings to incorporate absolute and relative positional information of tokens. However, when dealing with MSA, which represents stacked homologs, the structure is different. Each row of MSA corresponds to a distinct protein sequence, while each column represents the evolutionary trajectories of a specific amino acids (AAs) site.
To effectively capture the evolutionary patterns, recent approaches~\cite{jumper2021highly, rao2021msa, zhang2023enhancing} have employed decoupled axial attentions, which are designed to capture explicit co-evolutionary information along the rows (protein sequences) and columns (AAs sites).
However, these methods often suffer from low efficiency in capturing the information dynamics or fail to capture the evolutionary information adequately.

In light of this, we introduce a novel two-dimensional evolutionary positional encoding scheme, illustrated in Figure~\ref{fig:overall_frame}.
Given an MSA $\mathbf{M} \in \mathbb{A}^{N\times L}$, 
% where $N$ is the number of homogeneous sequences and $L$ is the number of AAs per sequence, 
% \cxy{This definition gets confused, RoPE is an un-learnable method.} 
we define a 2D positional id matrix $\mathbf{P}\in \mathbb{N}^{2\times N \times L}$, where the first positional id $\mathbf{P_0} \in \mathbb{N}^{1 \times N \times L} $ indicates the column position,  i.e., $P_0[i, \cdot] = \{0,1,\cdots, L\}$,  and 
the second positional id $\mathbf{P_1}$ indicates the row position, i.e., $P_1[j, \cdot] = \{0,1,\cdots, N\}$. The two positional ids are projected into two vectors 
% via learnable embedding tables, which are both 
added to the input token embeddings. 
% Practically, we adopt Rotary Positional Encoding (RoPE)~\cite{su2024roformer}. To fulfill the 2D nature of our evolutionary positional encoding, we implement
% the two-dimensional RoPE\footnote{https://kexue.fm/archives/8397} as our position embedding. 
We utilize the Rotary Positional Encoding (RoPE)~\cite{su2024roformer} technique, specifically adapting its two-dimensional variant\footnote{https://kexue.fm/archives/8397} to suit our 2D positional encoding requirements.

\begin{wrapfigure}{r}{0.58\textwidth}
  \centering
  \includegraphics[width=0.58\textwidth]{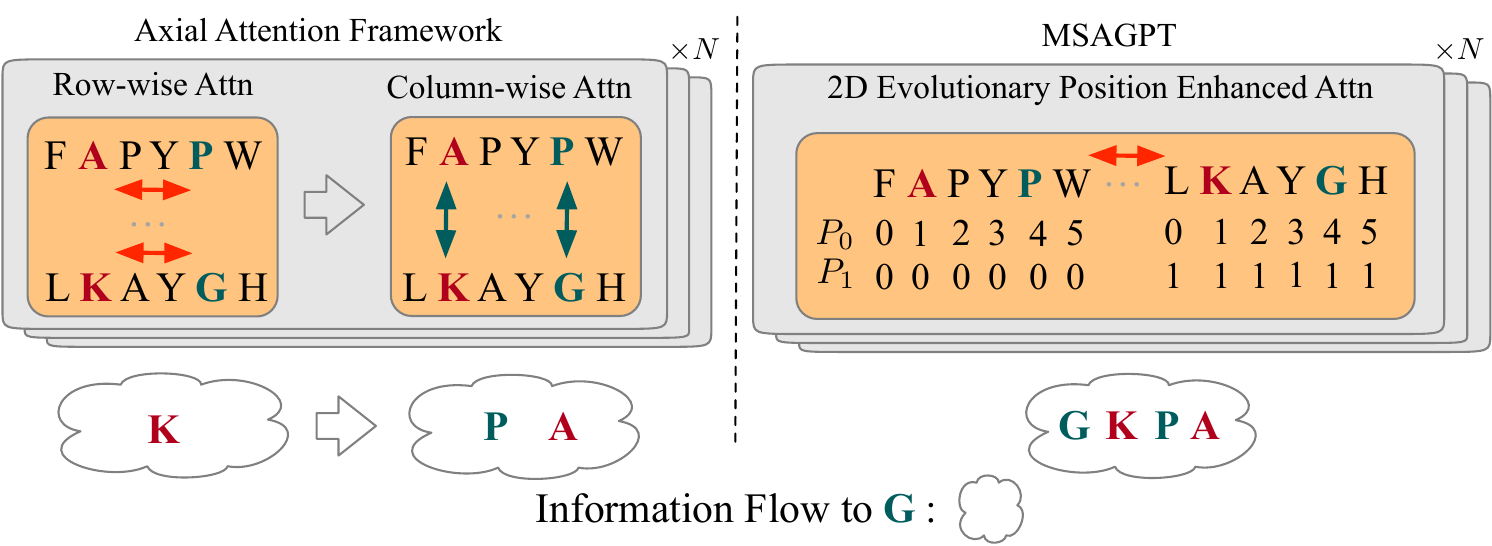}
  \caption{\label{fig:attn} Comparisons among the axial attention (exemplified by~\cite{rao2021msa}) and the one in \smodel in a single layer. Here we focus on the information aggregated to the AA ``G''. The 2D evolutionary position enhanced attention shows higher efficiency than the decoupled axial attentions with one-step aggregation to attain sufficient information.
 }
 \vspace{-10pt}
\end{wrapfigure}

\vpara{Comparison with Axial Attentions.} 
% \todo{The rational behind the 1D sequence decoding with 2D positional encoding. may be a figure to illustrate the differences between our information diffusion and axis attention.}
Considering the 2D positional id ($P_0$, $P_1$), the self-attention among AAs ($\alpha$, $\beta$) meets the following unit patterns, as illustrated in Figure~\ref{fig:attn}:

\noindent $\bullet$  $P^{\alpha}_0$ = $P^{\beta}_0$ \& $P^{\alpha}_1 \neq P^{\beta}_1$. Indicates $\alpha$ and $\beta$ reside in the same site across different protein sequences, such as the AA pair (A, K) and (P, G), 
% the alignment in the same site across various protein sequences, 
enabling column-wise self-attention that highlights evolutionary patterns conserved across sequences. 
\noindent $\bullet$  $P^{\alpha}_0 \neq P^{\beta}_0$ \&  $P^{\alpha}_1$ = $P^{\beta}_1$. Suggests $\alpha$ and $\beta$ are positioned in the same protein sequence but at different sites, such as the AA pair (A, P) and (K, G), facilitating row-wise self-attention that captures sequence-specific features.

\noindent $\bullet$  $P^{\alpha}_0 \neq P^{\beta}_0$ \&  $P^{\alpha}_1 \neq  P^{\beta}_1$. Denotes $\alpha$ and $\beta$ lack explicit correlation, such as the AA pair (A, G) and (P, K), may be serving as anchor nodes for complex co-evolutionary information diffusion.

Conceptually, the 2D positional encoding encapsulates the explicit row- and column-wise self-attention patterns with high efficacy. 
Moreover, it allows unrestricted information diffusion, that is, enabling any two amino acids to attend to one another. 
Such a framework facilitates unveiling complex co-evolutionary patterns, such as high-order correlations among AAs, that customized self-attentions might overlook.

\subsection{1D Zero-/Few-Shot MSA Decoding}
% Since the MSA input are 2D matrix, previous methods~\cite{rao2021msa, jumper2021highly, zhang2023enhancing, zhang2023unsupervisedly} directly manipulate the whole MSA using decoupled axis attention mechanisms. The interleaved attention calculation process not only fails to comprehensively characterize the co-evolutionary patterns,
% but also the 2D input itself is less friendly with the high-performance computing techniques adopted in the large language models like Flash Attention~\cite{dao2022flashattention, dao2023flashattention}.
% The 2D nature of MSA input is less friendly with the high-performance computing techniques adopted in the large language models like Flash Attention~\cite{dao2022flashattention, dao2023flashattention}. 
Leveraging the 2D evolutionary positional encoding, we further release the stacked MSA decoding task into the scalable 1D sequence generation framework, without compromising the integrity of co-evolutionary information.
Specifically, we convert the MSA $\mathbf{M} \in \mathbb{A}^{N\times L}$ into the flatted 1D sequence $\mathbf{M}^{f} \in \mathbb{A}^{1\times NL}$ along the row axis to ensure that we can generate the MSA sequentially during inference.
Similarly, the 2D positional id matrix $\mathbf{P}\in \mathbb{N}^{2\times N \times L}$ is reshaped into a flattened format, $\mathbf{P}^{f}\in \mathbb{N}^{1\times 2\times NL}$. 
This allows the model to conduct a simple auto-regressive generation process, as illustrated in Figure~\ref{fig:overall_frame}.

\vpara{Discussions.}
% To address the issue of capturing the intrinsic co-evolutionary information in MSA, we have proposed a 2D position encoding with 1D decoding framework. 
Admittedly, introducing 2D positional encoding introduces higher time complexity in comparison to conventional customized attention mechanisms (from $O(N^2L) + O(NL^2)$ to $O(N^2L^2)$). 
However, it is worth noting that the original stacked nature of MSA poses challenges for integrating it with acceleration techniques used in large language models, such as Flash Attention~\cite{dao2022flashattention, dao2023flashattention}. The 1D decoding framework, conversely, can be easily scaled to accommodate in-context learning frameworks, such as retrieval augmented generation, to further enhance the model's generation capability and expand its application range.
From a practical standpoint, the high parallelism of the 1D decoding framework demonstrates superior inference speed, benefiting from techniques like Flash Attention and KV-cache, while incurring negligible memory overhead compared to customized attention mechanisms. For further details, please refer to Appendix Section~\ref{subsec: efficiency}.

% facilitating the application of efficient transformer algorithms to accelerate the convergence of the pre-training process,

% Practically, the stacked nature of MSA presents challenges for integration with the high-performance computing techniques utilized in large language models, such as Flash Attention~\cite{dao2022flashattention, dao2023flashattention}. facilitating the application of efficient transformer algorithms to accelerate the convergence of the pre-training process,
% However, 
% To admit that, release the coevolution from customer attention to specifically designed position encodings brings the higher time complexity, from O(m) to O(n). However, the 1D encoding framework is easy compatible with hpc framework. What's more, fixble 1d framework the in-context learning framework such as retrievel-augmenta generation. which has a great potential to easy adopting to wide range of applications at ease. 

\section{The Training Pipeline of \model}
% In this section, we introduce the  pre-training objective, dataset, and training recipe. 
% Details please refer to the Appendix.
The training pipeline involves three successive stages: 
\textbf{Stage 1: MSA Generative Pre-Training} to obtain the base MSA generation model;  
% that comprehensively characterizes the intrinsic co-evolutionary patterns of MSAs;
\textbf{Stage 2: Rejective Fine-tuning (RFT)} to instruct the base model with high-quality MSAs via AF2 annotations, which can reduce generation hallucinations ;
\textbf{Stage 3: Reinforcement Learning from AlphaFold2 Feedback (RLAF)} to further enhance RFT model's capabilities based on the feedback of AF2. 
(See Appendix Section~\ref{sec: train_set} for training details.)
\subsection{Stage 1: MSA Generative Pre-Training}
\label{subsec: pre-train}
% Firstly, we obtain the initial base model through MSA generative pre-training. More specifically:
\vpara{Pre-Training Dataset.} 
We utilize the Uniclust30 MSA dataset from OpenProteinSet~\cite{ahdritz2024openproteinset}, which is processed through an all-against-all search on Uniclust30~\cite{mirdita2017uniclust} using HHblits~\cite{steinegger2019hh}. This results in approximately 16 million MSAs (See Appendix~\ref{subsec:app_data} for Details). 

\vpara{Pre-training Objective.} We adapt the language modeling objective~\cite{radford2019language} to the MSA generation task. The cross-entropy loss for modeling the intrinsic distribution of MSA $\mathbf{M}^{f} \in \mathbb{A}^{1\times NL}$ is defined as:
\beq{
    L_{\text{ce}}=\mathbb{E}_{\mathbf{M}^{f} }\left[\sum_{i=0}^{N\times L}-\log p(\mathbf{M}^f_{i} | \mathbf{M}^f_{<i}, \theta)\right]
}

\noindent where $\mathbf{M}^{f} \in \mathbb{A}^{1\times NL}$ is 1D flatted version of the input MSA and $\theta$ is the learned parameter.

\subsection{Stage 2: Rejective Fine-tuning (RFT)}
\label{subsec: sft}
% Although the base model is pre-trained on Uniclust30, 
% Due to that 
Noted that the pre-trained dataset inevitably contains noisy co-evolutionary patterns, such as large portions of deletions and insertions, which may mislead the base model to yield hallucinated cases, i.e., the linguistically reasonable but intrinsically unfaithful MSA. 
Thus we select highly-quality MSAs to further fine-tune the base model via a rejective sampling procedure based on the AF2-annotation.

\vpara{RFT Dataset.} We collect 120,780 protein sequences with structures from Protein Data Bank (PDB)~\cite{berman2000protein}. 
For the sequence $Q$, we search its MSA $M \in \mathbb{A}^{N\times L}$ from UniClust30~\cite{mirdita2017uniclust} with HHblits~\cite{steinegger2019hh}. 
Then we sample several MSA subsets $\mathbf{m}=\{m_1, m_2, ..., m_i\}$ with replacement, where $m_i \in \mathbb{A}^{n \times L}$ and $n\ll N$. 
To assure the information density of the sampled data, we filter out the MSA with depth $N$ fewer than $\lceil n\times i / 2 \rceil$.
Subsequently, we employ AF2 to score the sampled subset using the structure prediction accuracy $\mathbb{I}_{acc}(Q, m_i)$.  Then the RFT dataset $\mathcal{D}_{\text{RFT}}$ is defined as:
\beq{
    \mathcal{D}_{\text{RFT}} = \{(Q, m_i) | (\mathbb{I}_{acc}(Q, m_i)) > \theta_1 \cap (\mathbb{I}_{acc}(Q, m_i) - \mathbb{I}_{acc}(Q, -)) > \theta_2\}
}

\noindent where $\mathbb{I}_{acc}(Q, -)$ indicates the prediction accuracy without using MSAs. We set the sampling number $i=10$, the depth of sampled MSA $n=16$,  $\theta_1=0.9$, and $\theta_2=0.2$, which 
results in approximately 60k samples. 
The base model is fine-tuned on $\mathcal{D}_{\text{RFT}}$ with the same pre-training objective.

\subsection{Stage 3: Reinforcement Learning from AlphaFold2 Feedback (RLAF)}
% Learning from human feedback has been shown to be effective at aligning language models with human preferences~\cite{bubeck2023sparks, rafailov2023direct, ouyang2022training, christiano2017deep, schulman2017proximal}, which significantly improve the generation quality. 
% Existing methods adopt reinforcement learning from human feedback pipelines, implemented by the reinforcement algorithms. 
% Inspired by this, 
We further employ AF2 as the reward model to perform the Reinforcement Learning with AF2 Feedback (RLAF) using Direct Preference Optimization~\cite{rafailov2023direct} (DPO) to further guide the RFT model to decode meaningful structure-related MSA patterns that align with the preference of AF2. 

\vpara{RLAF Preference Dataset.} 
For each query $Q$ from the PDB, we use the RFT model to generate its MSA $M \in \mathbb{A}^{N\times L}$ in zero-shot manner. Then, we also sample several MSA subsets $\mathbf{m}=\{m_1, m_2, ..., m_i\}$ and obtain the preference dataset $\mathcal{D}_{\text{DPO}}=\left\{Q^{(k)}, m_{\text{w}}^{(k)}, m_{\text{l}}^{(k)}\right\}_{k=1}^{K}$ as follows,
\beq{
    \mathcal{D}_{\text{DPO}} = \{(Q, m_{\text{w}}, m_\text{l}) | \left(\mathbb{I}_{acc}(Q, m_{\text{w}}) - \mathbb{I}_{acc}(Q, m_\text{l})\right) > \theta_3\}
}

\noindent where we set the $\theta_3=0.3$, rendering the number of preference data $D_{\text{DPO}} = 11k$.
% The pipeline is illustrated in Appendix~\ref{sec:app_rlaf}.
% Then we filter those protein owns depth of MSA less than xxx.
% Given the dataset, we first use the pre-trained model generates multiple responses, i.e., multiple MSA set. Then we treat AlphaFold2 as the reward model to judge whether the generated virtual MSAs is high-quality (good) or not (bad). Then we construct the preference triplet \textit{(prompt, good response, bad response)} to further align the generation behaviours of our model to fulfilled the requirement of AlphaFold2, thus expecting to yielding better structure prediction accuracy.
% Guided by the AF2-preference signals, our model tends to generate the MSAs containing clean and sufficient co-evolutionary information rather than decaying to collapse point that just replicate partial sequence of the input query sequence.

% \cxy{Please give the loss function definition of DPO and AF2 rewards, such as the formulae in weekly reported PPT. }
\vpara{RLAF Training Objective.} The adapted DPO loss is defined as:
% to perform the Reinforcement Learning with AlphaFold2 Feedback (RLAF) training stage. 
% Because its computationally lightweight traits without the need for sampling from the LM during fine-tuning or performing significant hyperparameter tuning. 
% \beq{
%     L_{DPO}=\mathbb{E}_{\left(Q, m_{\text{w}}, m_{\text{l}}\right) \in D}\left[-\log \frac{\text{exp}\left(\beta \log \frac{\pi_{\theta} (m_{\text{w}}|Q)}{\pi_{\text{ref}} (m_{\text{w}}|Q)}\right)}{\text{exp}\left(\beta \log \frac{\pi_{\theta} (m_{\text{l}}|Q)}{\pi_{\text{ref}} (m_{\text{l}}|Q)}\right)} \right]
% }
\beq{
    L_{\text{DPO}}=\mathbb{E}_{\left(Q, m_{\text{w}}, m_{\text{l}}\right) \in \mathcal{D}_{\text{DPO}}}\left[-\log \sigma\left(\beta \log \frac{\pi_{\theta} (m_{\text{w}}|Q)}{\pi_{\text{ref}} (m_{\text{w}}|Q)}-\beta \log \frac{\pi_{\theta} (m_{\text{l}}|Q)}{\pi_{\text{ref}} (m_{\text{l}}|Q)}\right) \right]
}

\noindent where $\pi_{\theta}$ and $\pi_{\text{ref}}$ are initialized by the RFT model and $\pi_{\text{ref}}$ is frozen while $\pi_{\theta}$ is optimized.
% $\pi_{\theta}  (m_{\text{w}}|Q)$ implies the output probability of $m_{\text{w}}$ based on $Q$. 
During the RLAF training phase, we found that merely using the DPO loss led to training instability. Thus we adopt the pre-training loss $L_{\text{ce}}$ for the chosen answer $m_w$ as a regularization term with the coefficient factor $\lambda$ in the total loss to mitigate this issue. 
The total loss $L=L_{\text{DPO}} + \lambda L_{\text{CE}}$,  $\lambda=0.1$. Another critical coefficient $\beta$, which measures the penalty intensity for incorrect answers is set to $\beta=0.1$.
% More hyper-parameter studies refer to the Appendix Section .

\section{Experiments}
\label{sec:exp}

\begin{table}
	\newcolumntype{?}{!{\vrule width 1pt}}
	\newcolumntype{C}{>{\centering\arraybackslash}p{2.3em}}
	\renewcommand\tabcolsep{3.5pt} 
	\caption{
		\label{tb:natural} The performance of structure prediction on three natural MSA-scarce benchmarks. avg. Depth represents the average depth of searched MSA across all query sequences. Compared with the base model, the RFT and DPO models achieve higher TM-Score while with lower pLDDT values. (See Appendix Table~\ref{tb:natural-sup} for more results.)
	}
        % \tiny
        \small
        % \scriptsize
	% \footnotesize
	% \scriptsize
	%\centering 
	\renewcommand\arraystretch{1.0}
	\begin{tabular}{@{~}l?@{~}*{1}{CC?}*{1}{CC?}*{1}{CC?}*{1}{CC?}*{1}{CC?}*{1}{CC}@{~}}
		\toprule
            \multirow{3}{*}{\vspace{-0.3cm} Model}
            &\multicolumn{4}{c?}{\tabincell{c}{\textbf{CAMEO} \\ (avg. Depth = 8.5)} }
            &\multicolumn{4}{c?}{\tabincell{c}{\textbf{CASP} \\ (avg. Depth = 4.6)} }
            &\multicolumn{4}{c}{\tabincell{c}{\textbf{PDB} \\ (avg. Depth = 2.6)} }
            \\
            \cmidrule{2-7} \cmidrule{8-13} 
		% &\multicolumn{2}{c?}{\textbf{CAMEO (8.8)}}
            &\multicolumn{2}{c?}{\textbf{Zero-Shot}}
		&\multicolumn{2}{c?}{\textbf{Few-Shot}} 
		&\multicolumn{2}{c?}{\textbf{Zero-Shot} }
            &\multicolumn{2}{c?}{\textbf{Few-Shot} }
		&\multicolumn{2}{c?}{\textbf{Zero-Shot} } 
		&\multicolumn{2}{c}{\textbf{Few-Shot} }
		\\
		\cmidrule{2-7} \cmidrule{8-13} 
		& {pLDDT} & {TM} & {pLDDT} & {TM} & {pLDDT} & {TM} & {pLDDT} & {TM} & {pLDDT} & {TM} & {pLDDT} & {TM}   \\
		% xTFold
  %       & &  & - & - & - & -
		%   \\	
  %       AF2 Single
  %       & & 37.4 & - & - & - & -
  %       \\
		\midrule
        AF2 MSA
        &63.8 & 55.4 & 77.4 & 71.4 & 44.0 & 32.6 & 54.2 & 44.1 & 55.2 & 45.6 & 61.0 & 52.3 \\
  		EvoDiff
		& 68.0 &59.1 & 70.1 & 60.1 & 45.0&29.5 & 50.6 & 38.1 &54.8& 45.2 &59.1&47.2  \\	
		MSA-Aug.
		& 67.7& 59.2 & 77.4 & 72.1 & 56.8 & 36.6 & 63.4 & 46.3 & 61.9 & 49.8 & 66.0 & 55.3   \\	
		EvoGen
		& 66.1& 60.3& 78.6 & 75.3 & 48.2 & 38.4& 55.1 & 48.5 & 57.6 & 49.5& 62.8 & 55.4   \\	
		\midrule
  %       + \model-150M
		% &-   & 36.6 & -  & 47.5 & -  & 64.1  \\	
        \model
		& \textbf{70.8}& 61.4& \textbf{80.8} & 75.2 & \textbf{59.0} &39.8 & \textbf{65.4} & 51.0 & \textbf{68.6} & 53.4& \textbf{71.3} & 59.6    \\	
        + \textbf{RFT}
		& 68.0 & 60.5 & 79.8 & 76.4 & 56.8 & 40.2& 64.0 & 53.6 & 66.8 & 53.4& 70.3 & \textbf{60.1}    \\		
        + \textbf{DPO}
		& \tabincell{c}{68.9 \\ \textcolor{red}{(+3.1)}}  
  & \tabincell{c}{\textbf{62.7} \\ \textcolor{red}{(+2.4)}} 
  & \tabincell{c}{80.2  \\ \textcolor{red}{(+2.2)}}  
  & \tabincell{c}{\textbf{76.7}  \\ \textcolor{red}{(+1.4)}}
  & \tabincell{c}{54.2  \\ \textcolor{red}{(+2.2)}} 
  & \tabincell{c}{\textbf{43.7}   \\ \textcolor{red}{(+5.3)}}   
  & \tabincell{c}{62.7  \\ \textcolor{red}{(+2.0)}} 
  & \tabincell{c}{\textbf{57.0}   \\ \textcolor{red}{(+8.5)}}
  & \tabincell{c}{64.5  \\ \textcolor{red}{(+6.7)}} 
  & \tabincell{c}{\textbf{53.6}   \\ \textcolor{red}{(+3.8)}} 
  & \tabincell{c}{68.0  \\ \textcolor{red}{(+5.3)}} 
  & \tabincell{c}{59.7   \\ \textcolor{red}{(+4.7)}} 
  \\	
		\bottomrule
	\end{tabular}
	
\end{table}

\begin{table}
    \centering
    \begin{minipage}{0.50\textwidth}
	\centering
    \newcolumntype{?}{!{\vrule width 1pt}}
    \newcolumntype{C}{>{\centering\arraybackslash}p{2em}}
    \caption{
        \label{tb:arti} Zero-shot evaluation on artificial MSA-scarce benchmark (GDT stands for GDT-TS).
    }
    \footnotesize
    \centering 
    \renewcommand\arraystretch{1.0}
    \begin{tabular}{@{~}l?@{~}*{1}{CCCCCC}@{~}}
        \toprule
        \textbf{Model} & \textbf{pTM} & \textbf{pLDDT} & \textbf{TM} & \textbf{GDT} & \textbf{LDDT}\\
        \midrule
        AF2 MSA
        & 28.9 & 69.6 & 42.0 & 62.4 & 67.0 \\
        EvoDiff
        & 34.7 & 70.8 & 44.0 & 64.1 & 69.9\\
        MSA-Aug.
        & 33.2 & 74.3 & 45.8 & 65.9 & 71.7 \\
        EvoGen
        & 31.7 & 71.6 & 44.6 & 64.6 & 70.5 \\
        \midrule
        MSAGPT & \textbf{37.9} & \textbf{80.4} & 50.7 & 70.7 & 76.1 \\
        +\textbf{RFT \& DPO} & \textbf{37.9} & 79.7 & \textbf{51.6} & \textbf{71.4} & \textbf{76.9} \\
        \bottomrule
    \end{tabular}
    \end{minipage}\hfill
    % \hspace{-0.1in}
    \begin{minipage}{0.45\textwidth}
  \centering
    \newcolumntype{?}{!{\vrule width 1pt}}
    \newcolumntype{C}{>{\centering\arraybackslash}p{2.5em}}
    \renewcommand\tabcolsep{3.5pt} 
    \caption{
        \label{tb: selection} 
        Evaluation of selection methods.
        % Performance comparison between non-selection and pLDDT-selection models. 
    }
    \footnotesize
    % \scriptsize
    %\centering 
    \renewcommand\arraystretch{1.0}
\begin{tabular}{@{~}l?@{~}*{1}{C?}*{1}{C?}*{1}{C}@{~}}
		\toprule
        \multirow{1}{*}{Model}
		% &\multicolumn{2}{c?}{\textbf{CAMEO (8.8)}}
        &\multicolumn{1}{c?}{\textbf{CAMEO}}
		&\multicolumn{1}{c?}{\textbf{CASP}} 
		&\multicolumn{1}{c}{\textbf{PDB}}
		\\
            % \cmidrule{2-4}
            \midrule
            \model-DPO
		& 76.7 & 57.0 &  59.7   \\	
        + \textbf{pLDDT Selection}
		& \textbf{77.5} & \textbf{57.6} & \textbf{60.5}  \\		
		\bottomrule
	\end{tabular}

\hspace{-0.1in}
  \centering
    \newcolumntype{?}{!{\vrule width 1pt}}
    \newcolumntype{C}{>{\centering\arraybackslash}p{2.5em}}
    \renewcommand\tabcolsep{3.5pt} 
    \caption{
        \label{tb: other} 
        Evaluation on transfer learning.
        % Performance comparison between with or without virtual MSA generated by \smodel on four protein tasks. ACC is short for Accuracy. 
    }
    \footnotesize
    % \scriptsize
    %\centering 
    \renewcommand\arraystretch{1.0}
\begin{tabular}{@{~}l?@{~}*{1}{C?}*{1}{C?}*{1}{C?}*{1}{C}@{~}}
		\toprule
        \multirow{1}{*}{Model}
        &\multicolumn{1}{c?}{\textbf{CtP}}
		&\multicolumn{1}{c?}{\textbf{SsP}} 
		&\multicolumn{1}{c?}{\textbf{LocP}}
            &\multicolumn{1}{c}{\textbf{MIB}}
		\\
            \midrule
            w/o Virtual MSA
		& 11.6 & 66.5 &  \textbf{58.3} & 57.5   \\	
         \textbf{w/ Virtual MSA}
		& \textbf{13.1} & \textbf{69.0} &  56.4 & \textbf{60.3}  \\		
		\bottomrule
	\end{tabular}
    \end{minipage}
\end{table}

\subsection{Setup}
% \todo{1. Add large Natural MSA-scare benchmark; 2. Adding SFT \& DPO exps. Ablation study: a. SFT gap \& no gap, b. multi-rounds sft. DPO, a. natural vs natural, natural vs gen. \& gen vs gen; 3. Does MSA perform better on other tasks.}
% \todo{provide details of MSA generation}

\vpara{Benchmarked Dataset.} 
% We employ CAMEO~\cite{haas2018continuous}, CASP14/15 as our test set, the well-known protein structure benchmarks covering a broad spectrum of biological protein families. For each protein sequence of these datasets, we search its MSA by UniClust30~\cite{mirdita2017uniclust} with HHblits~\cite{steinegger2019hh}.
We employ the datasets from CAMEO~\cite{haas2018continuous}, CASP14\&15, and PDB~\cite{berman2000protein}, which are esteemed benchmarks in protein structure analysis spanning a diverse array of biological protein families. 
For each protein sequence, we search its MSA on UniClust30 database~\cite{mirdita2017uniclust} using HHblits~\cite{steinegger2019hh}.
% Each protein sequence's MSA was prepared using HHblits~\cite{steinegger2019hh} to perform an exhaustive search for homologs within the UniClust30 database~\cite{mirdita2017uniclust}.
% As our primary goal is to tackle the challenging cases with inferior MSA, thus hindering the efficacy of canonical MSA-based PSP methods like AlphaFold2, we define two MSA-scarce benchmarks as follows:
Given our focus on addressing the challenge presented by cases with limited MSA information, we establish two specific benchmarks to represent the MSA-scarce conditions: 
% we build the benchmark to represent the real-world MSA-scarce conditions. More specifically, we identify 200 protein sequences with the number of searched MSA fewer than 20 (8 from cameo, 13 from CASP14\&15, 179 from PDB).

\ipara{Natural MSA-scarce Benchmark.} We identify 200 protein sequences with the number of searched MSA fewer than 20 (8 from cameo, 13 from CASP14\&15, 179 from PDB).

\ipara{Artificial MSA-scarce Benchmark.} We collect approximately 8k protein sequences based on the PDB released before 2024-01-22 without searching MSA. 
% We perform zero-shot evaluations merely using the query sequence.

% This benchmark presents realistic challenges to MSA-based PSP models for accurate protein structure predictions. 
All MSA from the test set are removed from the pre-train dataset.

\vpara{Baselines.}
To assess the performance of \model, we adopt AF2 as the benchmark MSA-based PSP algorithm.
% because of its leading performance in the structural biology area. 
For MSA generation baselines, we compare \textbf{\model}, its RFT-version and its DPO-version with two advanced MSA generation algorithms: 
\textbf{EvoDiff}~\cite{alamdari2023protein}, which utilizes the diffusion framework for controllable protein sequence generation. Specifically, we use the EvoDiff-MSA for the MSA generation,
\textbf{MSA-Augmentor}~\cite{zhang2023enhancing}, which utilizes a sequences-to-sequences pre-training architecture incorporating an encoder and a decoder based on the axial attention~\cite{rao2021msa} and
\textbf{EvoGen}~\cite{zhang2023unsupervisedly}, 
% utilizes a diffusion-based generative model framework~\cite{ho2020denoising, ramesh2022hierarchical} resorting to the guidance from AlphaFold2 to refine its MSA generation.
which employs a meta generative model framework with customized attention, leveraging guidance from AF2 to refine its MSA generation.
As PoET~\cite{truong2024poet} is designed for mutational scanning tasks, we don't take it as the baseline. 
Additionally, we include the reference model \textbf{AF2 MSA}, which utilizes all the searched natural MSA for prediction. 

\vpara{MSA Generation Pipeline.} 
Given that \smodel can perform flexible zero- or few-shot MSA generation to accommodate different levels of available evolutionary information, we define two generation settings to evaluate models' performances under varying conditions:

\ipara{Zero-Shot Generation.} MSA generation is conducted using only the query sequence as input, emphasizing the model's ability to infer necessary evolutionary patterns without additional contexts.

\ipara{Few-Shot Generation.}
% As MSA exhibits important co-evolution information that serves as valuable contexts, 
All the searched natural MSA are viewed as the prompt to inform the few-shot MSA generation process. 
% More specifically, 
% for each query sequence, we select the top-1 and top-5 searched MSAs ranked by the sequence identity to the query in the descending order as the input prompt, termed as the \textbf{1-shot} and \textbf{5-shot} scenarios, respectively. We also take all the searched MSAs as the prompt, termed as \textbf{All-shot}. 
Then the generated MSA, combined with the initial prompts, serves as augmented data for structure predictions.

\vpara{Evaluation Metric.} 
% We employ TM-Score, a widely-used metric for assessing the structural similarity of proteins (normalized to the value in (0,1]),
% % and pLDDT, to assess the quality of protein structure predictions. 
% where 1 indicates a perfect match between two structures. Each experiment is conducted with 3 independent runs and we report the average performance
We use several widely-used metrics to assess structural similarity between predicted structures and ground truth: TM-Score, GDT-TS, and LDDT. Additionally, we include pTM and pLDDT, the corresponding predicted metrics estimated by AF2. All metrics are normalized from 0 to 100 for comparison, with higher scores indicating higher confidence and usually a more accurate prediction (see Appendix Section \ref{sec:eval_set} for details). 

\begin{figure}
    \centering
    \begin{minipage}{0.70\textwidth}
	\centering
	\subfigure[Effects of MSA Depths]{\label{fig:dpo_length}
		\includegraphics[width=0.42\textwidth]{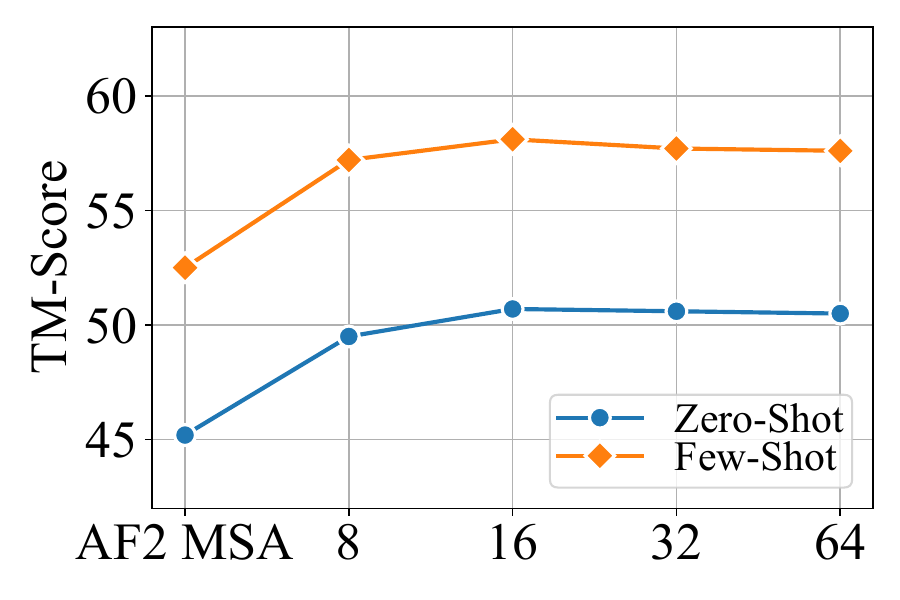}
	}
	\hspace{-0.1in}
	\subfigure[Effects of Selection Methods]{\label{fig:sel_vs}
		\includegraphics[width=0.55\textwidth]{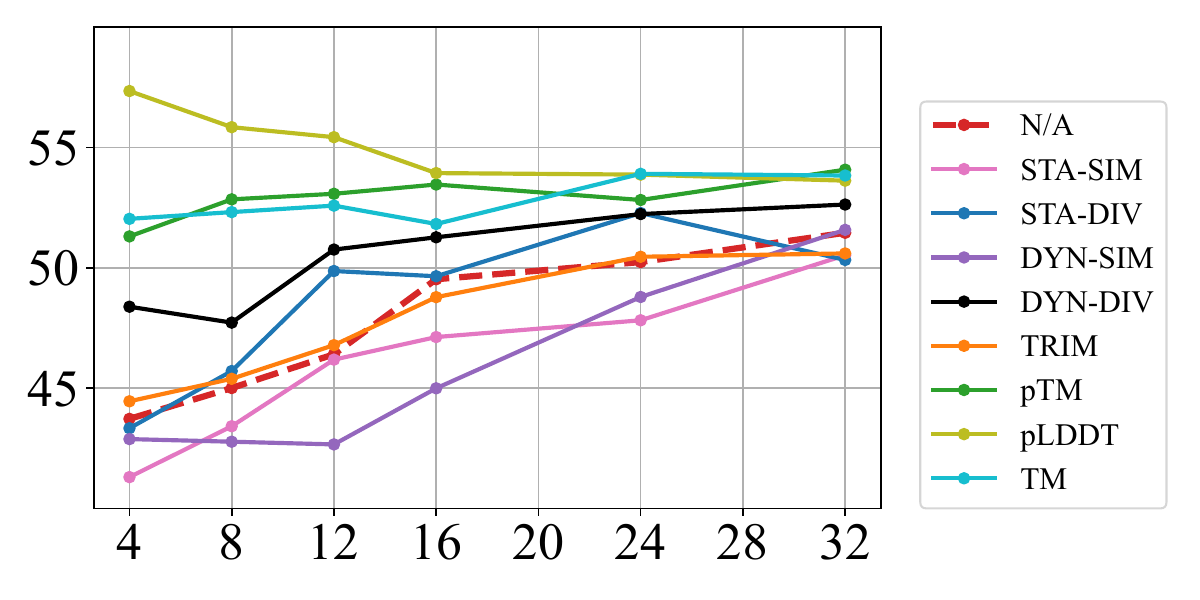}
	}
	\caption{\label{fig:selection_sec} The effect of different MSA depths and selection methods. The X-axis indicates the different MSA depths. The Y-axis represents the TM-Score. The dashed line denotes the non-selection baseline.
    }
    \end{minipage}\hfill
    \hspace{-0.1in}
    \begin{minipage}{0.28\textwidth}
      \centering
      \includegraphics[width=\textwidth]{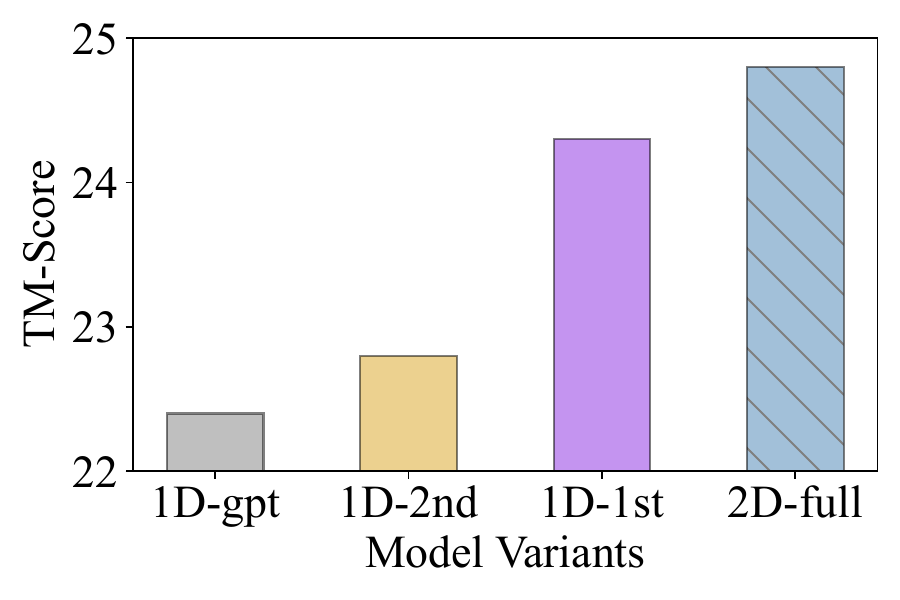}
      \caption{\label{fig:abla_vs} Ablation study with positional embedding variants.}
    \end{minipage}
\end{figure}

\subsection{\model's Virtual MSAs Reflect the Co-evolutionary Information}
% \label{subsec:arti}
% \vpara{Overall Results.} 
% Table~\ref{tb:arti} represents the overall performance of zero-shot/few-shot structure prediction among various baselines. 
% Among these, AF2 + Single performs the worst as it only use the query sequence as input without using MSA. 
% This underscores the importance of MSA quality for cutting-edge PSP algorithms. 
% Table~\ref{tb:natural} and ~\ref{tb:arti} showcase the comparative results in three datasets across different baselines. 
Table~\ref{tb:natural} and~\ref{tb:arti} showcase the comparative results in two benchmarks across different baselines. 
Notably, AF2 MSA, which relies solely on the limited searched MSA without incorporating virtual MSA, exhibits the worst performance. 
% As MSA-Augmenter and EvoGen are originally designed to generate virtual MSA under the guidance of the few-shot prompts, they even perform worse than AF2 + Single on the zero-shot generation scenario, which large limits its real-world applications on cases with inferior MSA.
% Interestingly, MSA-Augmenter and EvoGen, while specifically designed to generate virtual MSA with few-shot prompts, demonstrate inferior performance in the zero-shot generation scenario compared to AF2 + Single. This limitation significantly restricts their applicability in real-world situations where high-quality MSA is scarce.
Predictions enhanced with MSA generated by MSA-Augmentor or EvoGen surpass the performance of AF2 MSA. 
This underscores the critical role of high-quality MSA in enhancing the accuracy of cutting-edge PSP algorithms. 
% HoweverA key observation from the prompt MSA is their substantial proportion of gaps, indicating the low-quality issue with searched MSA.
% This low-quality input misleadingly impacts the baselines, leading to suboptimal MSA generation results.
% s is not their scarcity but rather their low quality. 
% \begin{wrapfigure}{r}{0.58\textwidth}
% 	\centering
% 	\subfigure[Zero-Shot]{\label{subfig:prompt}
% 		\includegraphics[width=0.28\textwidth]{figures/vs_base_full.pdf}
% 	}
% 	\hspace{-0.1in}
% 	\subfigure[Few-Shot]{\label{subfig:full}
% 		\includegraphics[width=0.28\textwidth]{figures/vs_full_final.pdf}
% 	}
% 	\caption{\label{fig:head_fig} The overall TM-Score comparisons between \model-DPO with AF2 MSA on Zero- and Few-Shot MSA generation scenarios in three benchmarks.
%     The solid line represents 95\% confidence interval. 
%     % Comparisons are conducted based on the 5-shot generation.
%     % X-axis represents TM-Score of \model-DPO, and the y-axis represents TM-Score of AF2 + prompt (a) and AF2 + Full (b). Notably, AF2 + Full 
%     }
% \end{wrapfigure}
% \vspace{6pt}
% our proposed method achieves the best results, significantly outperforming the AF2 MSA by a considerable margin. 
% This significant improvement not only demonstrates the superior accuracy and effectiveness of our MSA generation approach but also its robustness in handling cases with noisy or low-quality MSA. 
Overall, \smodel surpasses other advanced baselines by a large margin on both benchmarks, specifically achieving +1.4\% improvement on CAMEO, +8.5\% on CASP, and +4.7\% on PDB, as measured by TM-Score, on the natural MSA-scarce benchmark. This significant improvement demonstrates not only the superior accuracy and effectiveness of \smodel but also its robustness in handling cases with noisy or low-quality MSA.
Moreover, compared with the base model, the RFT and DPO models achieve higher golden metric scores, that is, GDT, LDDT, and TM-Score, but with a lower predictive score, that is, the value of pTM and pLDDT. This discrepancy might arise from the presence of highly confident (according to pTM and pLDDT) but lower-scored decoys (according to TM-Score), as observed in ~\cite{zhang2023unsupervisedly}, indicating that aligning with the preference dataset, which is filtered based on TM-Score, makes the model more inclined to generate truly informative MSA rather than hallucinated ones.
% Compared with the base model, the RFT and DPO models achieve higher TM-Score while with lower pLDDT values. Noted that we filter out the RFT and preference dataset based on the TM-Score, this result discrepancy indicates that there may exist some highly confident (according to pLDDT) but much lower-scored decoys (according to TM-Score), as observed in ~\cite{zhang2023unsupervisedly}.  
% This highlights the superiority of our proposed \smodel framework with the 2D evolutionary positional encoding mechanism and 1D flexible MSA decoding paradigm. 
% A fine-grained comparison is depicted in Figure~\ref{fig:head_fig},
% It’s evident that our generative virtual MSA effectively improve results for 84.6\% of protein sequences (over AF2 + Prompt). Remarkably, nearly half (44.8\%) of the generated MSA even outperform the real searched MSA. This emphasizes the potential of our proposed \smodel framework in unraveling co-evolutionary patterns within bio-sequences.
% Figure~\ref{fig:head_fig} reveals that 

Statistically, \smodel effectively improves the prediction accuracy for 91.0\% and 88.9\% of protein sequences with limited MSA when compared to AF2 MSA on Zero-Shot and Few-shot scenarios, respectively. 
% Remarkably, nearly half (\todo{44.8}\%) of the generated MSA outperform their real-world searched counterparts (All-shot). 
This significant finding highlights the potential of our \smodel framework to uncover and leverage co-evolutionary patterns within biosequences.
Notably, we also discuss the scenario with abundant natural MSA in the Appendix Section~\ref{subsec:msa_ab}.

\subsection{Rethinking the MSA Selection Strategy}
\label{subsec:select}
We further study the effect of different depths of virtual MSA, as shown in Figure~\ref{fig:dpo_length}. We observe a trend where the relative improvement in structure prediction accuracy decreases as the depth of virtual MSA increases. The accuracy based on MSA with 64 MSA sequences even underperforms those based on only 16 or 32 sequences.
% We observe that the relative improvement is deceased with the increment of the generated MSA depths. Surprisingly, the structure prediction accuracy with 64 homogeneous sequences of MSA matches or even underperforms that with 16 or 32 homogeneous sequences.
% The findings suggest a potential trade-off between MSA depth and the quality of co-evolutionary information captured. 
We hypothesize that increasing the number of virtual MSA beyond a certain threshold may introduce a dilution effect, where the density of valuable co-evolutionary signals is compromised by the inclusion of the hallucinated generation noise. 
% This phenomenon leads to a decrease in the marginal utility of adding more sequences to the MSA.
% We assume that as we generate more homogeneous sequences, the brought density of co-evolution information may be diluted by the hallucinate generation noise, which results in the diminishing marginal efficiency. 
To alleviate this, 
% we further discuss the MSA selection strategy about how to filter out those noisy sequence and remain the high-quality sequence that is beneficial to the structure predictions in Section~\ref{subsec:select}.  
we explore MSA selection strategies for filtering out low-quality, noise-inducing sequences while retaining those that contribute positively to the accuracy of structure predictions, as illustrated in Figure~\ref{fig:sel_vs} (See Appendix Section~\ref{sec:sel_app} for details). 
\vpara{1D Sequence Similarity or Diversity Measure.} 
We first arrange MSA by their similarity to the query sequence in descending order.
% , ranking MSA by their similarity to the query sequence from high to low. 
The results reveal that prioritizing MSA based on their high similarity to the query, termed as \textit{static similarity (STA-SIM)}, does not improve prediction accuracy compared to the non-selection approach (N/A). 
On the contrary, the \textit{static diversity (STA-DIV)} strategy, which favors MSA with lower similarity rankings, slightly outperforms the baseline, highlighting the importance of sequence diversity in enhancing MSA quality.
% Conversely, our \textit{static diversity (STA-DIV)} strategy, which selects the bottom-ranking sequences, marginally surpasses the baseline, underscoring the value of sequence diversity in MSA.
% \begin{wrapfigure}{r}{0.4\textwidth}
%   \centering
%   \includegraphics[width=0.40\textwidth]{figures/selection-curve-total.pdf}
%   \caption{\label{fig:sel_vs} Comparisons among different selection methods. The dashed red line indicates using all sequences at a given depth, while solid lines represent subsets selected from 48 sequences using specific strategies. 
%  % Curves smoothed by Exponential Moving Average (EMA) with alpha=0.3.
%  }
% \end{wrapfigure}
% \begin{wrapfigure}{r}{0.30\textwidth}
% 	\centering
% 	\includegraphics[width=0.29\textwidth]{figures/diff_depth.pdf}
% 	\caption{\label{fig:dpo_length} 
%  The effect of different MSA depth. AF2 + Pro. is short for AF2 + Prompt. 8, 16, 32, 64 indicate the different depths of generated MSA by \model-DPO.
%  }
% \end{wrapfigure}
% To delve into an depth insight, we employ the dynamic approach, 
Moreover, we employ the dynamic approach, initially selecting the most (or least) similar MSA to the query sequence and progressively incorporating additional MSA based on their average similarity to the cumulatively selected set, termed as \textit{dynamic similarity (DYN-SIM)} and \textit{dynamic diversity (DYN-DIV)}.

% \begin{wrapfigure}{r}{0.68\textwidth}
% 	\centering
% 	\subfigure[Effects of MSA Depths]{\label{fig:dpo_length}
% 		\includegraphics[width=0.28\textwidth]{figures/diff_length}
% 	}
% 	\hspace{-0.1in}
% 	\subfigure[Effects of Selection Methods]{\label{fig:sel_vs}
% 		\includegraphics[width=0.38\textwidth]{figures/selection-curve-total}
% 	}
% 	\caption{\label{fig:selection_sec} The effect of different MSA depths and selection methods. The X-axis indicates the different MSA depths. The Y-axis represents the TM-Score. The dashed line denotes the non-selection baseline.
%     }
% \end{wrapfigure}

% \vspace{5pt}
% This methodology, referred to as \textit{dynamic similarity (DYN-SIM)} and \textit{dynamic diversity (DYN-DIV)}, 
The results further confirm the advantage of fostering diversity within MSA
% suggesting that sequence diversity should be carefully considered when selecting MSA, 
rather than selecting only the sequences with high similarities to the query sequence.
% rather than solely in relation to the target sequence.
% initially selecting only the MSA sequence most / least similar to the query sequence and iteratively adding MSA based on their average sequence identity with \textbf{the entire growing selected set}. This method—the \textit{dynamic similarity (DYN-SIM)} and \textit{dynamic diversity (DYN-DIV)} strategies—confirmed the superiority of diversity, with \textit{DYN-DIV} notably enhancing MSA quality, suggesting that diversity should be evaluated within the entire set rather than solely in relation to the target sequence.
% \begin{wraptable}{r}{7cm}
%   \centering
%     \newcolumntype{?}{!{\vrule width 1pt}}
%     \newcolumntype{C}{>{\centering\arraybackslash}p{2.5em}}
%     \renewcommand\tabcolsep{3.5pt} 
%     \caption{
%         \label{tb: selection} Performance comparison between non-selection and pLDDT-selection models. 
%     }
%     \footnotesize
%     % \scriptsize
%     %\centering 
%     \renewcommand\arraystretch{1.0}
% \begin{tabular}{@{~}l?@{~}*{1}{C?}*{1}{C?}*{1}{C}@{~}}
% 		\toprule
%         \multirow{2}{*}{\vspace{-0.3cm} Model}
% 		% &\multicolumn{2}{c?}{\textbf{CAMEO (8.8)}}
%         &\multicolumn{1}{c?}{\textbf{CAMEO}}
% 		&\multicolumn{1}{c?}{\textbf{CASP}} 
% 		&\multicolumn{1}{c}{\textbf{PDB}}
% 		\\
%             \cmidrule{2-4}
% 		& TM & TM & TM   \\
%             \midrule
%             \model-DPO
% 		& 76.7 & 57.0 &  59.7   \\	
%         + \textbf{pLDDT Selection}
% 		& \textbf{77.5} & \textbf{57.6} & \textbf{60.5}  \\		
% 		\bottomrule
% 	\end{tabular}
% \end{wraptable}
We also inspect the effectiveness of the widely-adopted MSA \textit{trimming (TRIM)} strategy~\cite{zhang2023unsupervisedly}, 
% which emphasizes selecting MSA sequences closely resembling the query while ensuring their average similarity with other chosen MSA remains below 90\%. 
% The MSA \textit{trimming (TRIM)} strategy~\cite{zhang2023unsupervisedly}
which yields a similar TM-Score to the non-selection baseline, undermining its efficacy in selecting MSA with high quality. 
% but providing a means to simulate the quality of non-selected generated MSA.
% This strategy involves first filtering out MSA sequences with coverage less than 50\%, or sequence identity with the query sequence greater than 90\% or less than 20\%. Then it iteratively selects the MSA with the highest sequence identity to the query sequence and also an average sequence identity less than 90\% with all the currently selected MSA.

% \begin{figure}[t]
% 	\centering
% 	\includegraphics[width=0.49\textwidth]{figures/selection-curve-total.pdf}
% 	% \vspace{-10pt}
% 	\caption{\label{fig:sel_vs} \textbf{The TM-Score curves across different selection methods.} \textmd{The dashed red line indicates using all sequences at a given depth, marked as the reference model, while solid lines represent subsets selected from 48 sequences using specific strategies. 
%  % Curves smoothed by Exponential Moving Average (EMA) with alpha=0.3.
%  }}
% 	% \vspace{-10pt}
% \end{figure}
\vpara{3D Structure Affinity Measure.}
We assume that the generated sequence with high quality should exhibit structural congruity with the query sequence, thereby emitting strong co-evolutionary signals. 
To validate this, we rank sequences within MSA by their predicted tertiary structures according to the pTM, a predicted TM score~\cite{jumper2021highly}, pLDDT, and TM-Score, from highest to lowest.
% as depicted in Figure~\ref{fig:sel_vs}.
% Compared with 1D sequence metric, 
These approaches, especially when guided by the pLDDT score, consistently select high-quality MSA, evidenced by the enhanced TM-Score. 
% particularly for MSA of smaller depths.  
We compare the non-selection methods (N/A) and pLDDT selection methods on the three benchmarked datasets on few-shot generation scenarios in Table~\ref{tb: selection}. This confirms our hypothesis that structural similarity plays a crucial role in effective MSA selections.

\subsection{Transfer Learning of \smodel}
Since protein structures largely dictate their functions, the virtual MSA, enhancing structure prediction, should similarly benefit other protein tasks. To validate this, we focus on two protein structural tasks: Contact Prediction (CtP) and Secondary Structural Prediction (SsP) and two protein functional tasks: Localization Prediction (LocP) and Metal Ion Binding (MIB)~\cite{chen2024xtrimopglm}. We sample 1,000 sequences from each benchmark and conduct 5-fold cross-validation (See Appendix Section~\ref{subsec:trans} for details). 

\vpara{Results.} Table~\ref{tb: other} demonstrates that incorporating MSA from \smodel consistently surpasses merely using the single sequence on most tasks. Yet, it achieves inferior performance on the LocP task, which agrees with the observation~\cite{li2024feature} that protein language models may not present scaling behavior on several protein functional or property prediction tasks.  
Nevertheless, the results show the great potential of \smodel to contribute to a wide range of tasks with generated MSA. We are motivated to explore additional transfer tasks to assess MSAGPT's utility across various domains further.

\subsection{Ablation Study}
To understand the effect of various positional encoding strategies on capturing co-evolutionary patterns, we design four model variants: 
% to understand the effect of various positional encoding strategies on capturing co-evolutionary patterns:
\textbf{1D\_gpt}: Adopts the standard GPT positional encoding;
\textbf{1D\_2nd}: Utilizes only the second-dimensional of the 2D evolutionary positional encoding mechanism;
\textbf{1D\_1st}: Utilizes the first-dimensional positional encoding;
\textbf{2D\_full}: Implements the 2D evolutionary positional encoding mechanism (See Appendix Section~\ref{sec:eval_set} for details). 

\vpara{Results.} Figure~\ref{fig:abla_vs} showcases the TM-score distribution across different model variants. The 1D\_gpt exhibits the lowest performance, attributed to its simplistic approach of
% \begin{wrapfigure}{r}{0.30\textwidth}
%   \centering
%   \includegraphics[width=0.30\textwidth]{figures/abla_tms}
%   \caption{\label{fig:abla_vs} Ablation study with positional encoding variants.}
% \end{wrapfigure}
treating the MSA as a concatenation of homologous sequences, thereby failing to discern any co-evolutionary patterns.
% \begin{figure}[t]
% 	\centering
% 	\subfigure[TM-Score]{\label{subfig:abla_tms}
% 		\includegraphics[width=0.23\textwidth]{figures/abla_tms.pdf}
% 	}
% 	\hspace{-0.1in}
% 	\subfigure[pLDDT]{\label{subfig:abla_plddt}
% 		\includegraphics[width=0.23\textwidth]{figures/abla_plddt.pdf}
% 	}
% 	% \vspace{-10pt}
% 	\caption{\label{fig:abla_vs} \textbf{The TM-Score and pLDDT distributions among different model variants.}}
% 	% \vspace{-10pt}
% \end{figure}
Both the 1D\_1st and 1D\_2nd demonstrate significant improvement over 1D\_gpt, by explicitly encoding column- or row-wise relationships within the MSA, respectively. 
Notably, the performance of 1D\_1st is better than that of 1D\_2nd, suggesting that column-wise covariance patterns play a more crucial role in structural predictions than row-wise patterns. 
This aligns with the understanding that the permutation of sequence order does not alter the covariance information among residue sites~\cite{rao2021msa}.
Remarkably, the 2D\_full variant, which incorporates the proposed 2D evolutionary positional encoding, outperforms all other models, which underscores its effectiveness in capturing the intricate evolutionary information present in MSA.

\section{Limitations}
\label{sec:limit}
In this section, we discuss some limitations that should be resolved in future work.

\vpara{Scaling behavior of \model.}
While we have showcased the effectiveness of \smodel in generating informative virtual MSA, it is important to note that our pre-training was conducted with a model containing 2.8 billion parameters. The performance and behavior of \smodel, when scaled concerning dataset size, model size, and total compute resources, remain unknown.

% \vpara{General scenarios with sufficient homologs.}
% The primary objective of this paper is to improve the accuracy of protein structure prediction in cases where there is a scarcity of homologous sequences. However, whether we can further enhance the accuracy in scenarios where there are already sufficient MSA available, augmented by virtual MSA, remains an open question.

\vpara{Transfer Learning on a wide range of tasks.}
While we have demonstrated the transferability of \smodel on several tasks, including protein structure prediction and protein function prediction, its performance on a broader range of tasks remains an open question.
The ability of a model to transfer its learned knowledge and adapt to new tasks is a critical aspect of transfer learning. While \smodel has shown promising results on the tasks it was evaluated on, it is important to assess its performance on a more diverse set of tasks spanning various domains and problem types.

\section{Border Impact}
\label{sec:impact}
The aim of this paper is to improve the accuracy of protein structure prediction in cases with limited homologous sequences. The generated MSA also shows great potential to transfer to other protein-related tasks. By leveraging the information encoded in the generated MSAs, it is possible to enhance the performance of various protein-related tasks beyond structure prediction. However, the generative MSA may be misused to contaminate the high-quality nature MSA databases. Thus, it is necessary to train a classifier to distinguish the real from \model-generated MSA.

\section{Conclusion}
\label{sec:con}
% The paper introduce \model, a neural prompting protein structure prediction prediction via MSA generative pre-training model, to perform \textit{de novo} MSA generation for prompting the protein tertiary structure prediction in contexts where limited co-evolutionary information is available.
% To comprehensive characterize the co-evoluationary patterns within the MSA, such that it can generate high-quality MSA. \model proposes 2D Evolutionary Positional Encoding scheme with 1D Zero-/Few-Shot MSA Decoding mechanisms.
% The experical experiments on both the artificial or natural MSA-scarce benchmark have already validated \model's robustness and efficacy in generating MSA under zero-shot or few-shot scenarios flexibly.
% In  the future, we plan to adapt our model to more structure biology domain to enhance the taks which is highly resort to the co-evolutionary information in MSA.

This paper introduces \model, a novel approach that prompts protein structure prediction via MSA generative pre-training, to enable accurate protein structure predictions in situations where co-evolutionary information is scarce. 
To meticulously characterize the co-evolutionary patterns within MSA, 
% \smodel aims to generate high-quality MSA essential for precise structure prediction. Central to \model's methodology are 
\smodel designs two innovative techniques: the 2D Evolutionary Positional Encoding scheme and the 1D Zero-/Few-Shot MSA Decoding mechanisms. 
The post-alignment learning from AlphaFold2 feedback further enhances the quality of MSA generation. 
% These features collectively enhance the model's ability to capture and utilize evolutionary signals within protein sequences.
Empirical experiments conducted on a variety of benchmarks have demonstrated \model's robustness and effectiveness.
In the future, we plan to apply \smodel to broader areas, particularly for tasks that heavily rely on co-evolutionary information, and investigate the aforementioned limitations.
% \smodel has also shown its great potential in benefiting other protein-related tasks that heavily rely on co-evolutionary information of MSA. In the future, we plan to explore such as the scaling behavior of MSA Gnereative pre-training with \model.
% \smodel has also shown flexible generation capability, underscoring its potential to revolutionize protein structure prediction methodologies.
% In the future, we plan to apply \smodel to broader areas, particularly for the tasks that heavily rely on co-evolutionary information of MSA.

\vpara{Acknowledgments.}
This work has been supported by the NSFC for Distinguished Young Scholar 62425601, New Cornerstone Science Foundation through the XPLORER PRIZE and Tsinghua University Initiative Scientific Research Program.
% \clearpage
% \bibliographystyle{ACM-Reference-Format}
\bibliographystyle{unsrt}
\balance
\bibliography{main}
\clearpage
\appendix
\part{Appendix}
\parttoc
% \renewcommand{\contentsname}{Appendix Contents}

% % Show only appendix in ToC
% \addtocontents{toc}{\protect\setcounter{tocdepth}{-1}}

\section{Training Settings and Hyper-parameter Studies.}
\label{sec: train_set}
\begin{figure}[h]
    \centering
    \includegraphics[width=\textwidth]{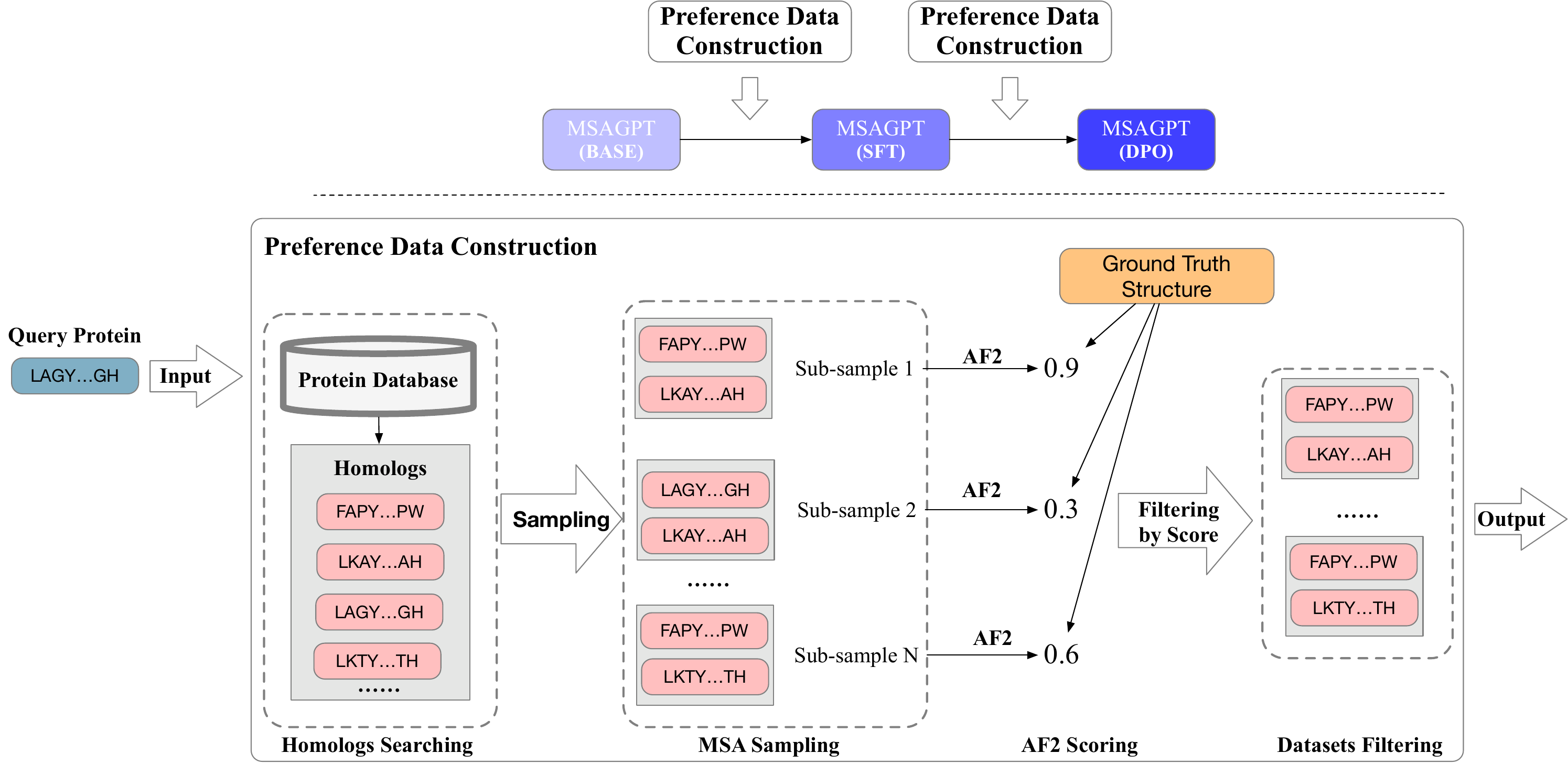}
    \caption{\label{fig:pipeline} \textbf{The overall training pipeline and the illustration of preference dataset construction process for SFT and DPO learning stages.}}
\end{figure}

The overall training pipeline is illustrated in Figure~\ref{fig:pipeline}.

\begin{table}
	\newcolumntype{?}{!{\vrule width 0.6pt}}
	\newcolumntype{C}{>{\centering\arraybackslash}p{1.8em}}
	\renewcommand\tabcolsep{3.5pt} 
	\caption{
		\label{tb:natural-sup} The performance (pTM, GDT, LDDT) of structure prediction on three natural MSA-scarce benchmarks. 
	}
        % \tiny
        % \small
        \scriptsize
	% \footnotesize
	% \scriptsize
	%\centering 
	\renewcommand\arraystretch{1.0}
	\begin{tabular}{@{~}l?@{~}*{1}{CCC?}*{1}{CCC?}*{1}{CCC?}*{1}{CCC?}*{1}{CCC?}*{1}{CCC}@{~}}
		\toprule
            \multirow{3}{*}{\vspace{-0.3cm} Model}
            &\multicolumn{6}{c?}{\tabincell{c}{\textbf{CAMEO} \\ (avg. Depth = 8.5)} }
            &\multicolumn{6}{c?}{\tabincell{c}{\textbf{CASP} \\ (avg. Depth = 4.6)} }
            &\multicolumn{6}{c}{\tabincell{c}{\textbf{PDB} \\ (avg. Depth = 2.6)} }
            \\
            \cmidrule{2-10} \cmidrule{11-19} 
		% &\multicolumn{2}{c?}{\textbf{CAMEO (8.8)}}
            &\multicolumn{3}{c?}{\textbf{Zero-Shot}}
		&\multicolumn{3}{c?}{\textbf{Few-Shot}} 
		&\multicolumn{3}{c?}{\textbf{Zero-Shot} }
            &\multicolumn{3}{c?}{\textbf{Few-Shot} }
		&\multicolumn{3}{c?}{\textbf{Zero-Shot} } 
		&\multicolumn{3}{c}{\textbf{Few-Shot} }
		\\
		\cmidrule{2-10} \cmidrule{11-19} 
		& {pTM} & {GDT} & {LDDT} & {pTM} & {GDT} & {LDDT} & {pTM} & {GDT} & {LDDT} & {pTM} & {GDT} & {LDDT} & {pTM} & {GDT} & {LDDT} & {pTM} & {GDT} & {LDDT}   \\
		% xTFold
  %       & &  & - & - & - & -
		%   \\	
  %       AF2 Single
  %       & & 37.4 & - & - & - & -
  %       \\
		\midrule
            AF2 MSA
            &41.3&60.8&66.7&58.2&75.5&80.3  &31.2&29.5&40.5&40.8&41.6&51.9   &40.3&43.7&56.6&46.2&50.4&62.8\\
  		EvoDiff
		&54.2&62.4&68.1&48.1&63.0&67.7  &40.2&27.3&38.4&44.1&35.6&46.7   &47.4&43.3&55.9&45.5&45.0&57.1\\	
		MSA-Aug.
		& 47.2&62.3&68.3&58.4&75.1&80.6  &39.1&33.2&42.7&44.3&43.2&51.9   &44.8&47.6&59.8&48.8&52.8&65.2\\	
		EvoGen
		&45.0&63.5&68.8&61.8&78.8&82.8   &36.0&34.8&44.9&42.8&45.1&54.6   &44.3&47.0&59.4&49.1&53.0&65.4\\	
		\midrule
  %       + \model-150M
		% &-   & 36.6 & -  & 47.5 & -  & 64.1  \\	
        \model
		&\textbf{51.4} &64.3 &\textbf{69.4} &\textbf{63.6} & 78.8&83.2   &40.0&36.3&45.5&47.7&47.8&56.0   &\textbf{48.6}&50.6&53.3&\textbf{53.7}&57.0&69.2 \\	
        + \textbf{RFT}
		& 48.1&63.5&68.1&62.4&79.0&83.1   &\textbf{41.2}&36.1&45.5&\textbf{49.8}&49.8&56.0    &\textbf{48.6}&50.5&62.4&53.6&57.2&69.0\\		
        + \textbf{DPO}
            &49.4&\textbf{64.4}&68.7&63.0&\textbf{79.4}&\textbf{83.5}   &40.0&\textbf{39.5}&\textbf{48.1}&\textbf{49.8}&\textbf{53.1}&\textbf{60.0}   &47.6&\textbf{50.9}&\textbf{63.2}&52.7&\textbf{57.8}&\textbf{69.5}\\
		\bottomrule
	\end{tabular}
\end{table}

\subsection{Pre-Training}
\label{subsec:app_data}
% The MSA dataset is parsed from UniClust30 (v2021\_03) clusters
% ~\cite{mirdita2017uniclust}. In order to obtain high-quality MSA, we screen MSA and sequences through the following process: we first screen out clusters with sequence lengths from 25 to 2000, and only retain sequences with the minimum identity of 30\% and the largest proportion of gap tokens no more than 10\%. And finally, clusters with more than 10 sequences are left. This results in ultimately leaving 1,663,204 MSAs as the training dataset.
% % The MSA dataset is generated for each UniRef50~\cite{suzek2007uniref} sequence by searching UniClust30
% % ~\cite{mirdita2017uniclust} with HHblits~\cite{steinegger2019hh}.
% % For each MSA, we further filter out those protein sequence with less than 0.3 sequence identity with the query sequence to guarantee the quality of the pre-trained datasets, which results in a dataset of xxx million MSAs and with xxx billion residues. The average depth and length of the MSAs is xxx and xxx, respectively.
% For detailed distribution, please refer to Appendix~\ref{sec:app_data}.
 To obtain high-quality MSA, 
% we screen MSA and sequences through the following process: 
we first screen out clusters with sequence lengths from 25 to 2000, and only retain sequences with the minimum identity of 30\% and the largest proportion of gap tokens no more than 10\%. 
The clusters with more than 10 sequences are left. 
As the MSA obtained by MSA search tools, inevitably contain noisy co-evolutionary patterns, such as large portions of deletions, insertions, and gaps. Many previous works aim to filter high-quality MSAs by clustering or ranking them based on predefined rules. One primary rule is to find MSA sequences most similar to the main sequence with fewer gaps, as these are more likely to represent informative co-evolutionary patterns. Following this idea, we sample MSA sequences using similarity-based weighted sampling, where sequences more similar to the query protein are more likely to be selected and ranked higher. Moreover, we also randomly shuffle the sequences in the selected MSA by a certain proportion to avoid injecting the order bias.
% \vpara{Pre-training Objective.} 

Regarding the backbone of \model,
we employ the standard transformer decoder framework~\cite{radford2019language, brown2020language} and 
% train two models with different size under the same setting, 
% \textbf{\model-150M}:
% 150 million parameters with 32 layers, 640 embedding size, and 20 attention heads;
% \textbf{\model-2.8B}:
% 2.8 billion parameters model with 36 layers, 2560 embedding size, and 40 attention heads.
train the model with 2.8 billion parameters owning 36 layers, 2560 embedding size, and 40 attention heads.
We employ batches of 48 MSAs with each MSA containing 12,288 residues. 
% Due to that the average length of MSA is xxx, one batch owns about $48 \times 12288 / 250 = 3,072$ protein sequences.  
We follow BF16 mixed-precision pre-training strategy.
We use AdamW~\cite{loshchilov2018decoupled} as our optimizer with $\beta_1=0.9$, $\beta_2=0.95$, $\text{eps}=10^{-8}$ and a learning rate of $1.2\times 10^{-4}$. 
We use a cosine learning rate schedule, with a warmup of the first 2.5\% steps, and decay the final learning rate down to 10\% of the peak learning rate.
We use a weight decay of 0.1 and gradient clipping of 1.0 without dropout.
% Each flatted MSA contains a fixed sequence length of 12,288 (We concatenate all protein sequences within MSA with a separator into a single document, and sample protein sequences from this document in such a way that there is virtually no padding during pre-training.). 
For the tokenization of the protein data, we use the residue-level tokenizer which is adopted in several PLMs~\cite{lin2023evolutionary,  chen2024xtrimopglm, nijkamp2023progen2}. 
% Except for the basic amino acid types, we add special
% tokens [MASK], [sMASK], and [gMASK] for model prediction. We also add special tokens
% <sop>, <eop>, <eos> for sequence separation (Cf. Table S8 for the full configurations)
To save the GPU memory and accelerate the pre-training process, we substitute the standard self-attention module with the Flash Attention-v1~\cite{dao2022flashattention} in each layer.
All models are trained on 24 A800 GPUs for 254k updates, consuming about 150 billion tokens. This process consumes approximately 150 billion tokens, requiring around 2.7 x $10^{18}$ floating point operations (FLOPs).

\begin{figure}[t]
    \centering
    \includegraphics[width=\textwidth]{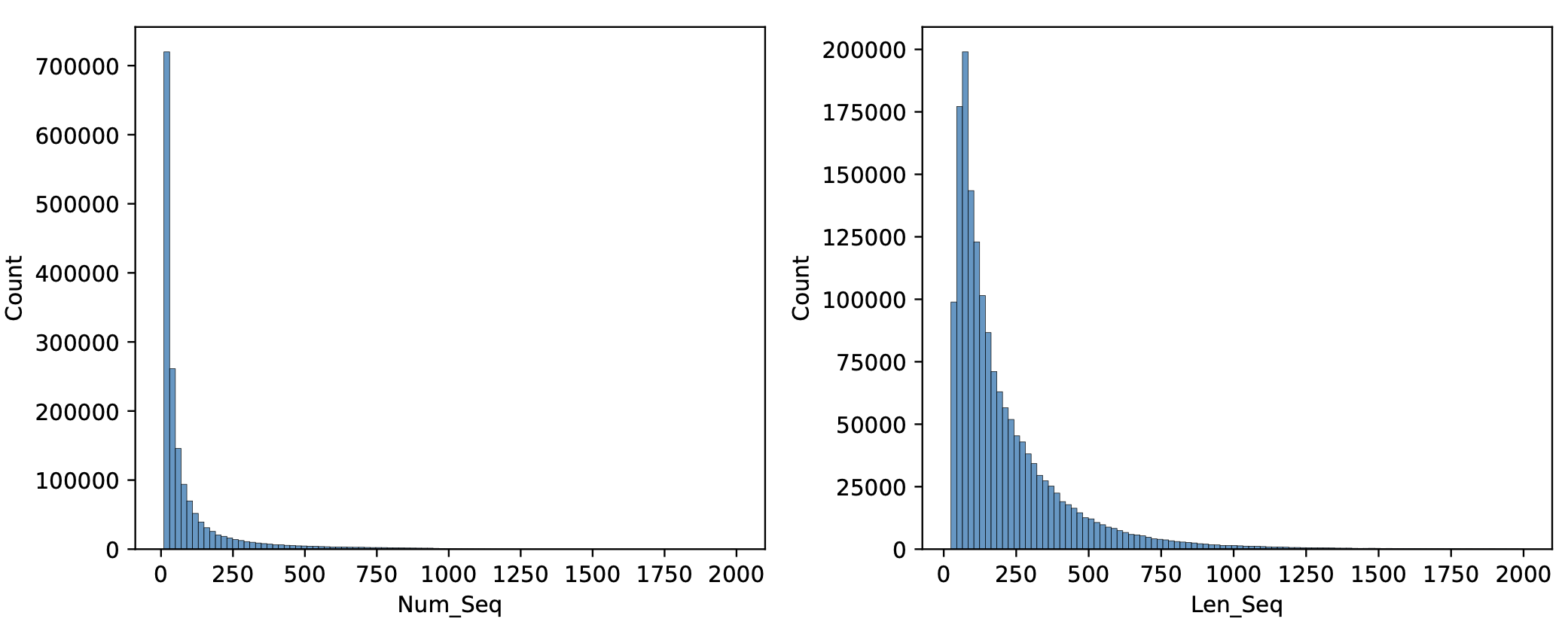}
    \caption{\label{fig:app_data} \textbf{The length and depth distribution of the pre-training dataset.}}
\end{figure}

\begin{figure}[t]
    \centering
    \includegraphics[width=0.8\textwidth]{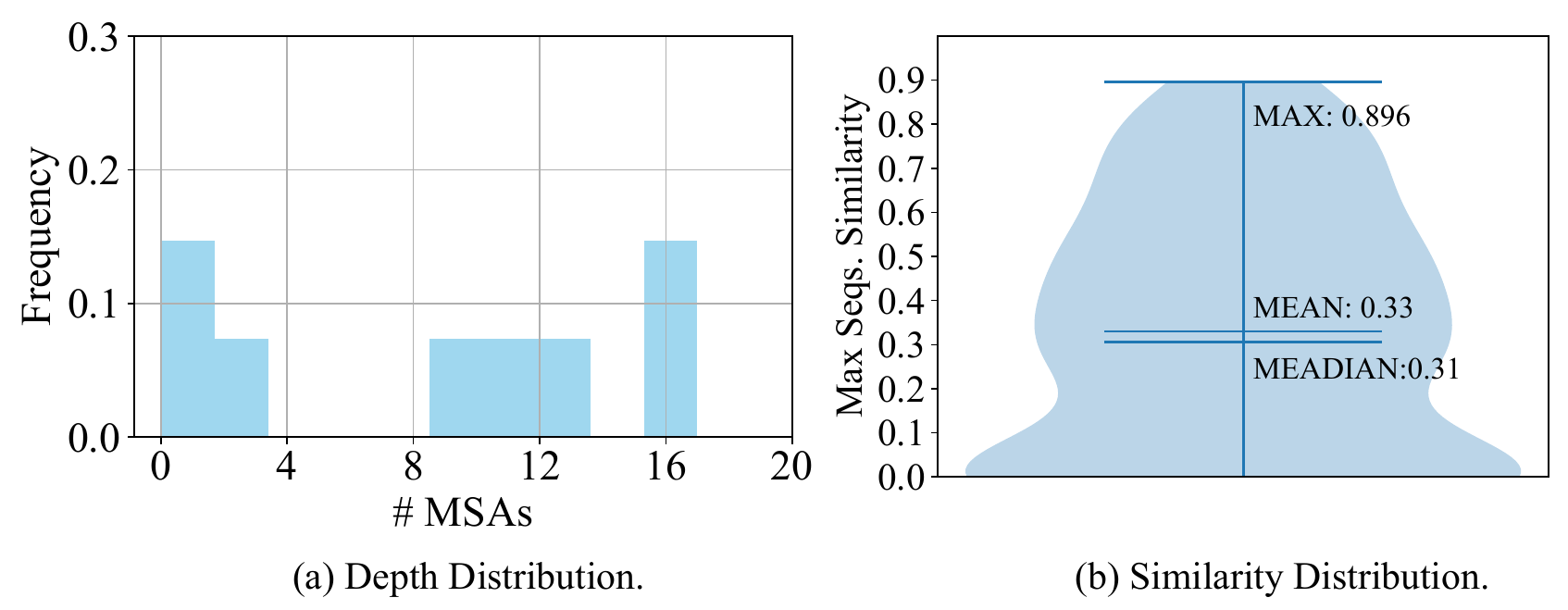}
    \caption{\label{fig:depth_sim} \textbf{(a) The distribution of MSA depth of benchmarked datasets and (b) the similarity distribution of sequences in the test set, as retrieved from the pre-training set using HHblits.}}
\end{figure}

\vpara{Pre-trained Dataset}
\label{sec:pre_data}
Figure~\ref{fig:app_data} illustrates the length and depth distribution of the pre-training dataset. Moreover, we implement a thorough filtering process to eliminate any potential data leakage. Specifically, we remove all MSAs of sequences in the test sets (CAMEO, CASP, and PDB) from the pre-training dataset. Furthermore, we ensure that any sequence in the pre-training set with a similarity greater than 0.9 to a sequence in the test set is excluded.
To validate this filtering process, we used the HHblits~\cite{steinegger2019hh} tool to retrieve sequences from the test set and calculate their maximal similarity distribution with sequences in the pre-training dataset. The results, illustrated in Figure~\ref{fig:depth_sim}(b) show that the maximum similarity is 0.89, confirming that there is no data leakage in the pre-training dataset.

\subsection{RFT}
We fine-tune the base model using the pre-training cross-entropy loss on $\mathcal{D}_\text{RFT}$ with training only one epoch. Specifically, 
we adopt the same experimental settings as that used in the pre-training stage, except for the learning rate of $1.0\times 10^{-5}$ by default.
Following the pre-training phase, the model undergoes a rejective fine-tuning process, which is more energy-efficient. This stage is executed on 8 x A800 GPUs for a single epoch for about two days.

% \begin{wraptable}{r}{7cm}
%   \centering
%     \newcolumntype{?}{!{\vrule width 1pt}}
%     \newcolumntype{C}{>{\centering\arraybackslash}p{2.5em}}
%     \renewcommand\tabcolsep{3.5pt} 
%     \caption{
%         \label{tb: RFT-thr} Performance comparison between different relative improvement threshold $\theta_2$ values. 
%     }
%     \footnotesize
%     % \scriptsize
%     %\centering 
%     \renewcommand\arraystretch{1.0}
% \begin{tabular}{@{~}l?@{~}*{1}{C?}*{1}{C?}*{1}{C}@{~}}
% 		\toprule
%         \multirow{1}{*}{\vspace{-0cm} Threshold $\theta_2$}
% 		% &\multicolumn{2}{c?}{\textbf{CAMEO (8.8)}}
%         &\multicolumn{1}{c?}{\textbf{0}}
% 		&\multicolumn{1}{c?}{\textbf{0.2}} 
% 		&\multicolumn{1}{c}{\textbf{0.5}}
% 		\\
%   %           \cmidrule{2-4}
% 		% & TM & TM & TM   \\
%             \midrule
%             \model + \textbf{RFT}
% 		& 61.2 & \textbf{62.5} &  61.3   \\	
%   %       + \textbf{pLDDT Selection}
% 		% &  &  &   \\		
% 		\bottomrule
% 	\end{tabular}
% \end{wraptable}

\vpara{RFT Dataset Filtering Threshold}
When curating the RFT dataset, we first sample multiple MSA subsets for each protein structure, and select high-quality MSA subsets based on the following criteria: (1) the absolute structure prediction accuracy using the MSA subset, as measured by TM-score, should be larger than $\theta_1$, and (2) the relative improvement of the prediction accuracy after using the MSA subset, as compared to single sequence prediction, should be larger than $\theta_2$. We set $\theta_1=0.9$, and experiment with different $\theta_2$ values, as shown in table~\ref{tb: RFT-thr}. The RFT model trained with dataset filtered by $\theta_2=0.2$ yields the best result, indicating that the relative information gain provided by MSA is a valuable indicator for curating high quality datasets for RFT. Moreover, $\theta_2=0.5$ results in a 20\% decrease in dataset size, leading to inferior RFT model performance, highlighting the necessity of an intricate balance between data quality and data volumn.

\subsection{RLAF}
We fine-tune the RFT model using the DPO algorithm on $\mathcal{D}_{\text{DPO}}$ with training only one epoch. Specifically, 
we adopt the batch size of 1 with each MSA subset containing a maximum of 16,384 residues.
We also use AdamW~\cite{loshchilov2018decoupled} with the learning rate of $1.0\times 10^{-6}$ by default. 
We linearly warmup the learning rate from 0 to $1.0\times 10^{-6}$  over the first 0.1\% steps.
This stage is also executed on 8 x A800 GPUs for a single epoch for about one day

\begin{figure}
	\centering
	\includegraphics[width=0.58\textwidth]{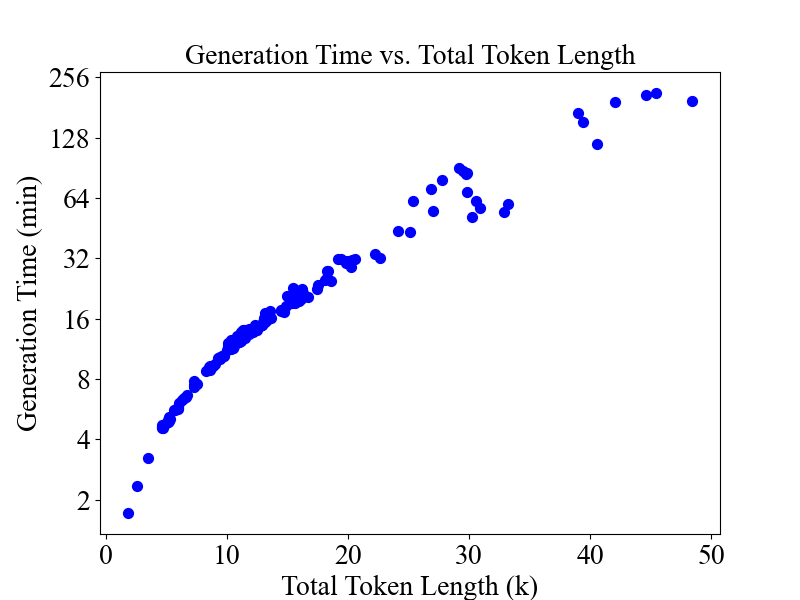}
	% \vspace{-10pt}
	\caption{\label{fig:scale} The correlation between total token length (the protein sequence length multiplied by the number of generated MSAs) and the inference time (minutes).
In most cases (total token length < 20K), the generation time of \smodel is lower than the AF2 search pipeline requiring more than 30 minutes.
The result shows \smodel can generate substantial sequence lengths within practical time, thus affirming its scalability and efficiency. 
}
\end{figure}

\begin{table}
  \centering
    \newcolumntype{?}{!{\vrule width 1pt}}
    \newcolumntype{C}{>{\centering\arraybackslash}p{2.5em}}
    \renewcommand\tabcolsep{3.5pt} 
    \caption{
        \label{tb: RLAF-thr} Performance comparison between different data source and filtering threshold values. 
    }
    \footnotesize
    % \scriptsize
    %\centering 
    \renewcommand\arraystretch{1.0}
\begin{tabular}{@{~}l?@{~}*{1}{C?}*{1}{C?}*{1}{C?}*{1}{C}@{~}}
		\toprule
        \multirow{1}{*}{\vspace{-0cm} Data Source}
		% &\multicolumn{2}{c?}{\textbf{CAMEO (8.8)}}
        &\multicolumn{1}{c?}{\textbf{nature(0.2)}}
        &\multicolumn{1}{c?}{\textbf{nature(0.3)}}
		&\multicolumn{1}{c?}{\textbf{generated(0.3)}} 
		&\multicolumn{1}{c}{\textbf{nature(0.3)+generated(0.4)}}
		\\
  %           \cmidrule{2-4}
		% & TM & TM & TM   \\
            \midrule
            \model + \textbf{RLAF}
		& 62.6 & \textbf{64.5} & \textit{63.5} &  62.7   \\	
  %       + \textbf{pLDDT Selection}
		% &  &  &   \\		
		\bottomrule
	\end{tabular}
\end{table}

\begin{table}
     \centering
    \newcolumntype{?}{!{\vrule width 1pt}}
    \newcolumntype{C}{>{\centering\arraybackslash}p{2.5em}}
    \renewcommand\tabcolsep{3.5pt} 
    \caption{
        \label{tb: RFT-thr} Performance comparison between different relative improvement threshold $\theta_2$ values. 
    }
    \footnotesize
    % \scriptsize
    %\centering 
    \renewcommand\arraystretch{1.0}
\begin{tabular}{@{~}l?@{~}*{1}{C?}*{1}{C?}*{1}{C}@{~}}
		\toprule
        \multirow{1}{*}{\vspace{-0cm} Threshold $\theta_2$}
		% &\multicolumn{2}{c?}{\textbf{CAMEO (8.8)}}
        &\multicolumn{1}{c?}{\textbf{0}}
		&\multicolumn{1}{c?}{\textbf{0.2}} 
		&\multicolumn{1}{c}{\textbf{0.5}}
		\\
  %           \cmidrule{2-4}
		% & TM & TM & TM   \\
            \midrule
            \model + \textbf{RFT}
		& 61.2 & \textbf{62.5} &  61.3   \\	
  %       + \textbf{pLDDT Selection}
		% &  &  &   \\		
		\bottomrule
	\end{tabular}
\end{table}

\vpara{RLAF Dataset.}
We conducted experiments with different data sources and filtering thresholds $\theta_3$—defined as the minimum relative improvement of the good case over the bad case in DPO data pairs—for the RLAF dataset, as detailed in Table~\ref{tb: RLAF-thr}. Utilizing only natural MSA subsets sampled from PDB, we found that higher $\theta_3$ values lead to improved model performance, suggesting a correlation between the disparity within data pairs and DPO effectiveness. Interestingly, the quality of MSA subsets generated by the RFT model surpasses that of natural MSA subsets at a $\theta_3$ of 0.2. However, the performance declines when natural MSAs are mixed with generated MSAs, compared to using a single data source during training. This indicates that maintaining distribution homogeneity is crucial for effective DPO alignment.
% We experiment with different data sources and filtering threshold $\theta_3$ values (minimum relative improvement of the good case over the bad case in the DPO data pairs) for the RLAF dataset, as shown in table~\ref{tb: RLAF-thr}. When using only natural MSA subsets sampled from PDB, a higher $\theta_3$ leads to better model performance, indicating that the performance gap within data pairs corresponds to data quality. Interestingly, when using MSA subsets generated by the RFT model, its quality exceeds that of natural MSA subsets at $\theta_3=0.2$. However, when mixing nature MSA with generated MSA, the model performance dropped compared to using only one data source, suggesting that the distribution homogeneity is necessary for effective DPO alignment.

\subsection{Inference Efficiency}
\label{subsec: efficiency}
Generally, it's vital to consider not just the immediate resource consumption during pre-training and post-alignment, but also the long-term utilization of these models. Once pre-trained, \smodel demonstrates significant efficiency, capable of generating protein sequences with up to 100,000 amino acids in under 8 hours. This efficiency underscores the model's value, especially when amortized over its application lifespan and subsequent fine-tunings for specific tasks.

Regarding the scalability of the \model. We present the inference time with different total lengths (measured by protein sequence length multiply the number of generated sequences.), as shown in Figure~\ref{fig:scale}.

% \section{The framework of RLAF}
% \label{sec:app_rlaf}
% Figure~\ref{fig:app_rlhf} illustrates the pipeline of reinforcement learning from AlphaFold2 feedback.

% \begin{figure}
%     \centering
%     \includegraphics[width=0.8\textwidth]{figures/msa_freq}
%     \caption{\label{fig:msa_dis} The distribution of MSA depth of benchmarked datasets.}
% \end{figure}

% \begin{figure}[t]
% 	\centering
% 	\subfigure[Depth Distributions.]{\label{subfig:msa_dis}
% 		\includegraphics[width=0.45\textwidth]{figures/msa_freq}
% 	}
% 	\hspace{-0.1in}
% 	\subfigure[Similarity Distributions.]{\label{subfig:max_sim}
% 		\includegraphics[width=0.40\textwidth]{figures/max_sim}
% 	}
% 	% \vspace{-10pt}
% 	\caption{\textbf{(a) The distribution of MSA depth of benchmarked datasets and (b) the similarity distribution of sequences in the test set, as retrieved from the pre-training set using HHblits.}}
% \end{figure}

The result showcases \model's ability to generate substantial sequence lengths within practical time frames, thus affirming its scalability and efficiency.

% \begin{wrapfigure}{r}{0.4\textwidth}
%   \centering
%   \includegraphics[width=0.40\textwidth]{figures/selection-curve-total.pdf}
%   \caption{\label{fig:sel_vs} Comparisons among different selection methods. The dashed red line indicates using all sequences at a given depth, while solid lines represent subsets selected from 48 sequences using specific strategies. 
%  % Curves smoothed by Exponential Moving Average (EMA) with alpha=0.3.
%  }
% \end{wrapfigure}

\begin{table}
        \centering
	\newcolumntype{?}{!{\vrule width 1pt}}
	\newcolumntype{C}{>{\centering\arraybackslash}p{2.3em}}
	\renewcommand\tabcolsep{3.5pt} 
	\caption{
		\label{tb:test} The paired Student's t-test between \smodel and other baselines on three benchmarks based on the TM-Score, where the p-value less than 0.05 indicates the result is said to be statistically significant.
	}
        % \tiny
        \small
        % \scriptsize
	% \footnotesize
	% \scriptsize
	%\centering 
	\renewcommand\arraystretch{1.0}
	\begin{tabular}{@{~}l?@{~}*{1}{C?}*{1}{C?}*{1}{C?}*{1}{C?}*{1}{C?}*{1}{C}@{~}}
		\toprule
            \multirow{2}{*}{\vspace{-0.3cm} Model}
            &\multicolumn{2}{c?}{\tabincell{c}{\textbf{CAMEO} \\ (avg. Depth = 8.5)} }
            &\multicolumn{2}{c?}{\tabincell{c}{\textbf{CASP} \\ (avg. Depth = 4.6)} }
            &\multicolumn{2}{c}{\tabincell{c}{\textbf{PDB} \\ (avg. Depth = 2.6)} }
            \\
            \cmidrule{2-7}
            % \midrule
		% &\multicolumn{2}{c?}{\textbf{CAMEO (8.8)}}
            &\multicolumn{1}{c?}{\textbf{Zero-Shot}}
		&\multicolumn{1}{c?}{\textbf{Few-Shot}} 
		&\multicolumn{1}{c?}{\textbf{Zero-Shot} }
            &\multicolumn{1}{c?}{\textbf{Few-Shot} }
		&\multicolumn{1}{c?}{\textbf{Zero-Shot} } 
		&\multicolumn{1}{c}{\textbf{Few-Shot} }
		\\
		% xTFold
  %       & &  & - & - & - & -
		%   \\	
  %       AF2 Single
  %       & & 37.4 & - & - & - & -
  %       \\
		\midrule
            AF2 MSA
            &0.014 & 0.023 & 0.008 & 0.007 & 5e-7 & 8e-9 \\
            EvoDiff
            &0.016 & 1e-5 & 5e-4 & 7e-5 & 0.012 & 1e-8 \\
		MSA-Aug. 
		& 0.023& 0.014& 0.044 & 0.015 & 6e-6 & 1e-7 \\
            EvoGen
            &0.038& 0.027 & 0.067 & 0.016 & 1e-8  &1e-9
		   \\	
		\midrule
  %       + \model-150M
		% &-   & 36.6 & -  & 47.5 & -  & 64.1  \\	
        \model
		& - &  - & - & - & -&-    \\		
		\bottomrule
	\end{tabular}
	
\end{table}

\section{Experimental Settings}
\label{sec:eval_set}
\subsection{Evaluation Details} 
We employ three golden metrics:
TM-Score, a widely-used metric for assessing the structural similarity between predicted structures and ground truth, LDDT, the local distance difference test score measures how well local interactions in a reference structure are conserved the protein model being assessed, 
GDT-TS, the global distance test to represent ``total score'', is a measure of similarity between two protein structures with known amino acid correspondences (e.g. identical amino acid sequences) but different tertiary structures, 
and two predicted metrics:
pLDDT, the predicted local distance difference test measuring the local confidence of  per-residue and
pTM, an integrated measure of how well AlphaFold2 has predicted the overall structure.
All metrics are normalized from 0 to 100 for comparison, with higher scores indicating higher confidence and usually a more accurate prediction.
% and pLDDT, to assess the quality of protein structure predictions. 
where 1 indicates a perfect match between two structures. Each run across 3 independent runs. For each run, we adopt the different temperatures (T $\in$ \{0.8, 1.0\}) along with different nucleus sampling factors (P $\in$ \{0.8, 1.0\}), experimenting with varying the number of generated MSAs in 8, 16, 32, and 64. The final performance is determined by first identifying the predicted structure with the highest accuracy across these different depths, and then averaging the results across the test set.

\subsection{Experiments on MSA-abundant Scenario}
\label{subsec:msa_ab}
We compare the results of query sequences with abundant natural MSAs to those with abundant natural MSAs augmented by MSAGPT's generated MSAs on CAMEO set. For this comparison, we sample 128, 256, and 512 sequences from both the natural MSAs and the generated MSAs, as shown in Table~\ref{tb:msa_ab}. These results indicate that the inclusion of generated MSAs has no significant effect on the performance in MSA-abundant conditions, which is consistent with previous findings that when more than 64 MSAs as input, AF2 predicts a ``converged'' structure.

\subsection{Setup of Transferability of \smodel to Other Tasks}
\label{subsec:trans}
We utilized the MSA Transformer~\cite{rao2021msa} as the backbone model with the task-specific heads. We finetune MSA transformer with the head with $\text{lr}=3e-5$ and ${\text{batchsize}= 16}$ on all experiments.
All the task benchmarks are obtained following the pipeline in ~\cite{chen2024xtrimopglm}. For each task, we sample 1000 protein sequences with the corresponding labels. Then we use \model-DPO to generate 32 virtual MSAs with the generation strategy T=0.8 and P=0.8.
Both setups are trained briefly (for one epoch) for 5-fold cross-validation as shown in Table~\ref{tb:cross}, and we report the average performance. 
% The ABR task involved classifying proteins based on the type of antibiotic they degrade. We compared the performance of two approaches: one using only a single protein sequence and another augmenting the single sequence with 16 MSAs generated by MSAGPT. 
\begin{table}
  \centering
    \newcolumntype{?}{!{\vrule width 1pt}}
    \newcolumntype{C}{>{\centering\arraybackslash}p{3em}}
    \renewcommand\tabcolsep{3.5pt} 
    \caption{
        \label{tb:cross} The results of 5-fold cross-validation performance between with or without virtual MSA generated by \smodel on four protein-related tasks. 
    }
    \footnotesize
    % \scriptsize
    %\centering 
    \renewcommand\arraystretch{1.0}
\begin{tabular}{@{~}l?@{~}*{1}{C?}*{1}{C?}*{1}{C?}*{1}{C?}*{1}{C?}*{1}{C}@{~}}
		\toprule
        \multirow{2}{*}{\vspace{-0.3cm} Model}
        &\multicolumn{1}{c?}{\textbf{1}}
		&\multicolumn{1}{c?}{\textbf{2}} 
		&\multicolumn{1}{c?}{\textbf{3}}
            &\multicolumn{1}{c?}{\textbf{4}}
		&\multicolumn{1}{c?}{\textbf{5}}
            &\multicolumn{1}{c}{\textbf{AVG}}
		\\
            \cmidrule{2-7}
		& Top (L/5) & ACC & ACC & ACC & ACC & -  \\
            \midrule
            w/o Virtual MSA (CtP)
		& 11.4 & 14.3 & 10.7 & 9.9 & 11.8 & 11.6   \\	
            \textbf{w/ Virtual MSA (CtP)}
		& 14.1 & 13.7 & 13.4 & 11.8 & 12.3 & \textbf{13.1}  \\	
            \midrule
            w/o Virtual MSA (SsP)
		& 67.7 & 65.8 & 64.0 & 68.9 & 66.2 & 66.5   \\	
            \textbf{w/ Virtual MSA (SsP)}
		& 70.5 & 67.8 & 67.5 &70.5 & 69.0 & \textbf{69.0}  \\	
            \midrule
            w/o Virtual MSA (LocP)
		& 56.0 & 64.5 & 48.0 & 59.0 &57.0 & \textbf{58.3}   \\	
            \textbf{w/ Virtual MSA (LocP)}
		& 47.0 & 58.5 & 53.5 & 64.0 & 59.0 & 56.4   \\	
            \midrule
            w/o Virtual MSA (MIB)
		&58.0 & 53.5 & 49.5 & 59.0 & 67.5 & 57.5   \\	
            \textbf{w/ Virtual MSA (MIB)}
		& 61.5 & 57.0 & 63.0 & 53.0 & 67.0 & \textbf{60.3}  \\	 
		\bottomrule
	\end{tabular}
\end{table}

\begin{table}
    \newcolumntype{?}{!{\vrule width 1pt}}
    \newcolumntype{C}{>{\centering\arraybackslash}p{5em}}
    \caption{
        \label{tb:msa_ab} Performance comparison in MSA-abundant scenarios across all 194 cases in the CAMEO datasets.
    }
    \footnotesize
    \centering 
    \renewcommand\arraystretch{1.0}
    \begin{tabular}{@{~}ll?@{~}*{1}{CCC}@{~}}
        \toprule
        \textbf{\# Natural MSA} & \textbf{\# Virtual MSA} & \textbf{TM} & \textbf{GDT-TS} \\
        \midrule
        0 & 0 & 39.1 & 28.5\\
        \midrule
        128 & 0 & 85.1 & 82.0\\
        128 & 128 & 85.3& 82.2\\
        \midrule
        256 & 0 & 86.2 & 83.8 \\
        256 & 256 & 86.0 & 83.5 \\
        \midrule
        512 & 0 & 86.5 & 84.2 \\
        512 & 512 & 86.4 & 84.0 \\
        \bottomrule
    \end{tabular}
    % \vspace{-20pt}
\end{table}

\subsection{Setup of Ablation Study}
Experiments are conducted based on models with 150 million parameter size encompassing 30 layers, 20 attention heads, 640 embedding dimensions. 
These models are trained across approximately 30 billion tokens, amounting to around 40k training steps, maintaining consistency in hyper-parameter settings with \model, except for variations in the positional encoding mechanism. The efficacy of each variant is assessed through zero-shot MSA generation on the CASP test set.

\section{Selection Strategy Details and pLDDT Evaluation}
\label{sec:sel_app}
\begin{wrapfigure}{r}{0.4\textwidth}
	\centering
	\includegraphics[width=0.4\textwidth]{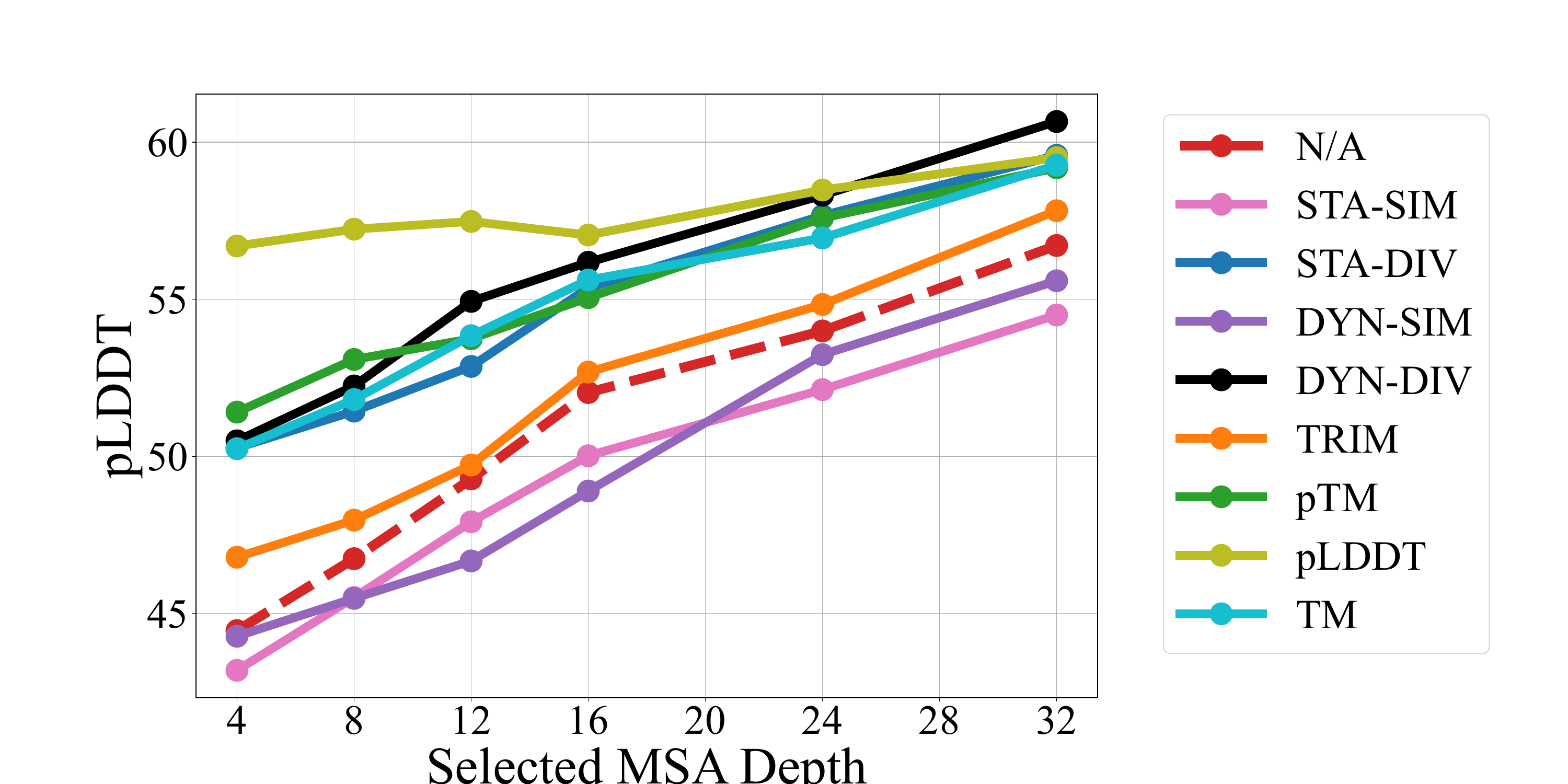}
	% \vspace{-10pt}
	\caption{\label{fig:sel_plddt} \textbf{The pLDDT curves across different selection methods.} \textmd{Dashed red line represents using all generated sequences of a given depth. Solid lines represent selecting a subset of a given depth from 48 generated sequences with a specific strategy. The curves are smoothed using the Exponential Moving Average with alpha=0.3.}}
	% \vspace{-10pt}
\end{wrapfigure}

To evaluate the effectiveness of different selection strategies, we extracted 4, 8, 12, 16, 24, and 32 sequences from 48 zero-shot generated MSA for each method and computed the median TM-scores (Figure~\ref{fig:sel_vs}) and pLDDT scores (Figure~\ref{fig:sel_plddt}) across 33 test cases. The strategies are detailed below.

\vpara{\textit{Static Similarity / Static Diversity} Strategy:}We select the top-k generated MSA with the highest / lowest sequence identity to the query sequence. Sequence identity is determined by the percentage of identical residues between the two sequences.

\vpara{\textit{Dynamic Similarity / Static Diversity} Strategy:}Starting with the MSA most / least similar to the query sequence, we sequentially incorporate MSA into the selected set based on the highest or lowest average sequence identity with all sequences already included, until reaching a total of k MSA.

\vpara{\textit{Trimming} Strategy:}Suggested by EvoGen, this method filters out MSA with less than 50\% coverage or sequence identity to the query sequence above 90\% or below 20\%. Subsequently, it iteratively selects the MSA with the closest sequence identity to the query and an average sequence identity below 90\% with all the chosen MSA.

\vpara{\textit{pTM / pLDDT / TM Score} Strategy:}For each generated MSA, we remove the gaps and predict its structure using AF2. The structures are then ranked based on the pTM score (as reported by AF2), the pLDDT score (as reported by AF2), or the TM score compared to the query sequence's ground truth structure (calculated by US Align), and the MSA corresponding to the top-k structures for each metric are selected accordingly.

\section{Protein Structure Prediction Analysis compared with natural MSA}
\label{sec:natural_case}
\begin{figure}
    \centering
    \includegraphics[width=0.9\linewidth]{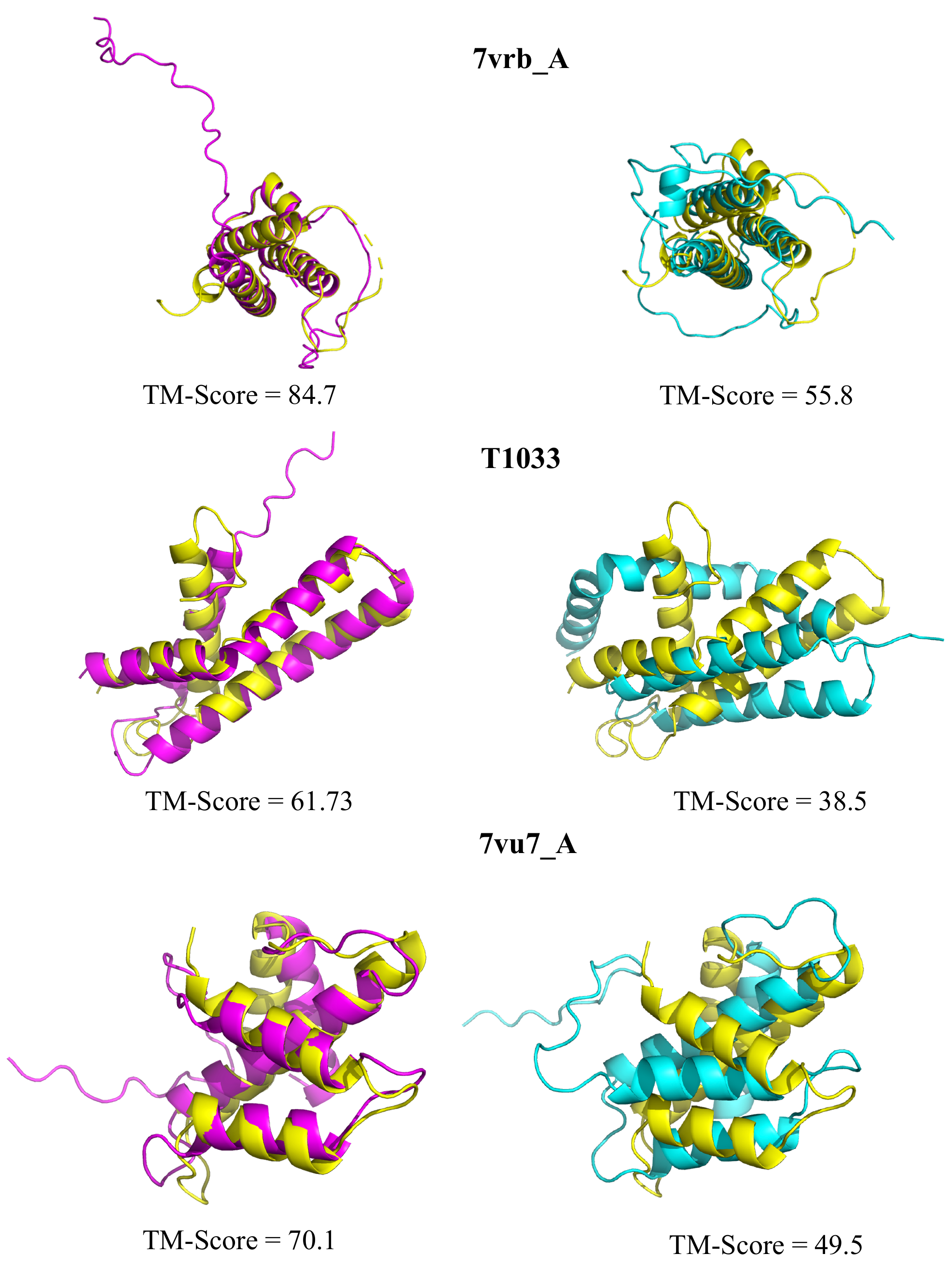}
    \caption{\label{fig:compare-baseline} Visualization of improved structure prediction compared with nature MSA
    \textmd{\textcolor{yellow}{Yellow}: Ground truth; 
    \textcolor{violet}{Pink}: Predictions based on MSA generated by \model; 
    \textcolor{cyan}{Blue}: Predictions from MSA generated by natural MSA.}}
\end{figure}

We present a detailed visual comparison of protein structures predicted by AlphaFold2 (AF2) utilizing MSA augmented by \model, against those predicted with natural MSA. This comparison, as depicted in Figure~\ref{fig:compare-baseline}, highlights the remarkable capability of \model in enhancing the accuracy of structure predictions to levels that closely rival, and in some cases surpass, those based on naturally derived MSA.

\begin{figure}
    \centering
    \includegraphics[width=0.8\linewidth]{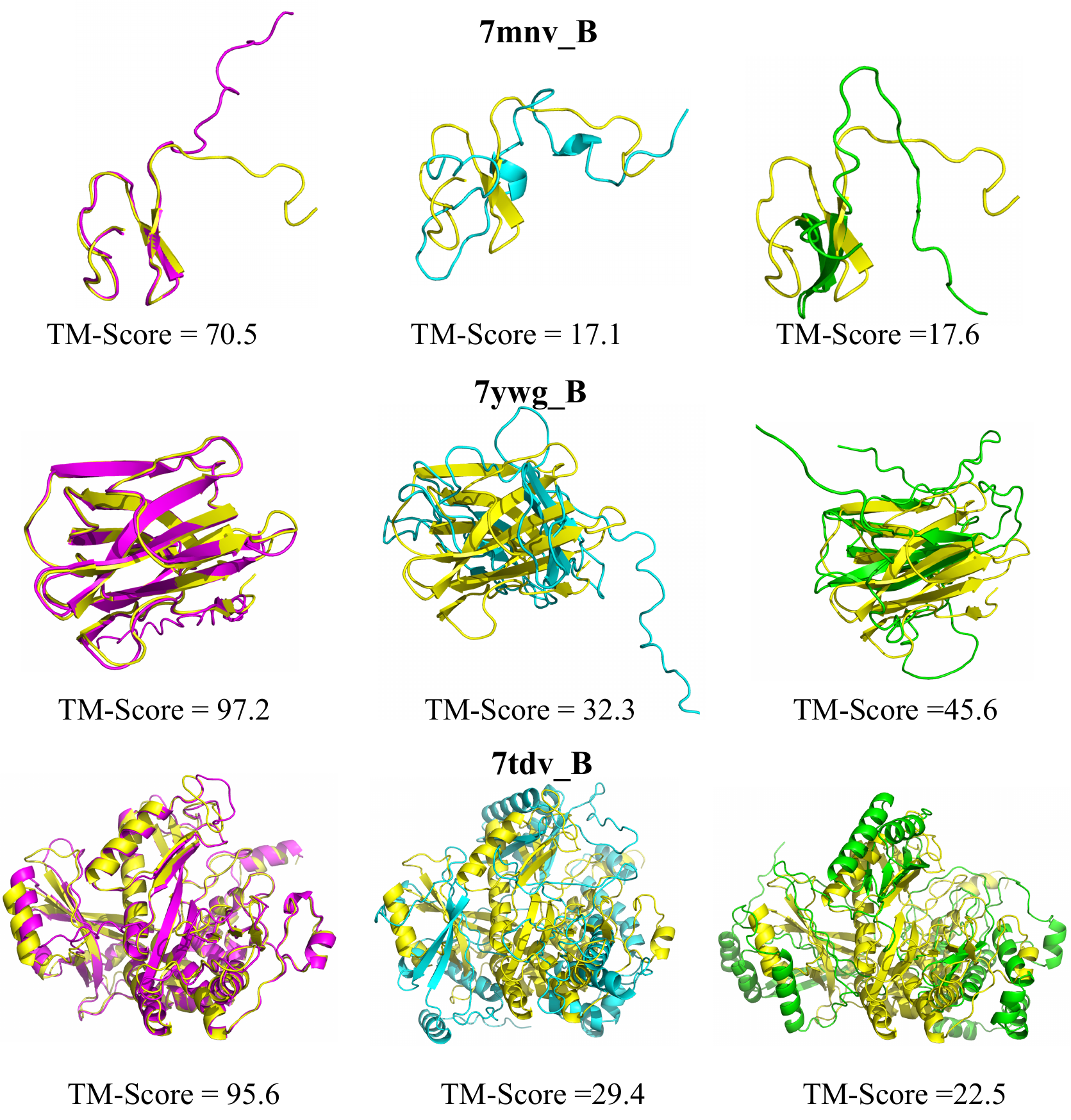}
    \caption{\label{fig:cases} 
    Visualization of improved structure prediction compared with baseline models. 
    \textmd{\textcolor{yellow}{Yellow}: Ground truth; 
    \textcolor{violet}{Pink}: Predictions based on MSA generated by \model; 
    \textcolor{cyan}{Blue}: Predictions from MSA generated by EvoGen;
    \textcolor{green}{Green}: Predictions utilizing MSA generated by MSA-Augmenter.}
    }
\vspace{-10pt}
\end{figure}

We delve into a visualized analysis of protein structures predicted using AlphaFold2 (AF2) with MSA augmented by our proposed model (\model), alongside those augmented by EvoGen and MSA-Augmenter. This comparison, visualized in Figure~\ref{fig:cases}, encompasses a spectrum of proteins, ranging from short sequences with relatively simple structures, like 7mnv\_B, to long sequences with complex configurations, such as 7tdv\_B. It includes proteins characterized by multiple beta sheets, exemplified by 7ywg\_B, as well as those rich in alpha helices, such as 7tdv\_B. Across these diverse cases, \smodel significantly surpasses both EvoGen and MSA-Augmenter, improving the TM score to a maximum of 0.9.

By detailed examination, we observe that while the MSA augmented by the baseline models assist AF2 in accurately predicting local structures and folds, they fall short in aligning the global composition and orientation with the ground truth structure, which is effectively addressed by MSA generated by \model. 
The local structures, which are generally more discernible from the spatial arrangements of adjacent amino acids, contrast with the global structures whose prediction relies heavily on comprehensively understanding the co-evolutionary information within MSA. 
These co-evolutionary patterns, indicating proximity in three-dimensional space through simultaneous mutations at multiple positions, are crucial for accurate global structure prediction.
These findings underscore \model's impressive capability to comprehensively capture and utilize co-evolutionary information, thereby significantly enhancing the accuracy of protein structure predictions. 
More visualization cases about the predictions based on MSA generated by \smodel and the predictions based on the natural MSA are illustrated in Appendix~\ref{sec:natural_case}.

\vspace{-10pt}

\section{Protein Structure Prediction Improvement after DPO}
\label{sec:app_dpo3b}
\begin{figure}[t]
	\centering
	\includegraphics[width=0.9\linewidth]{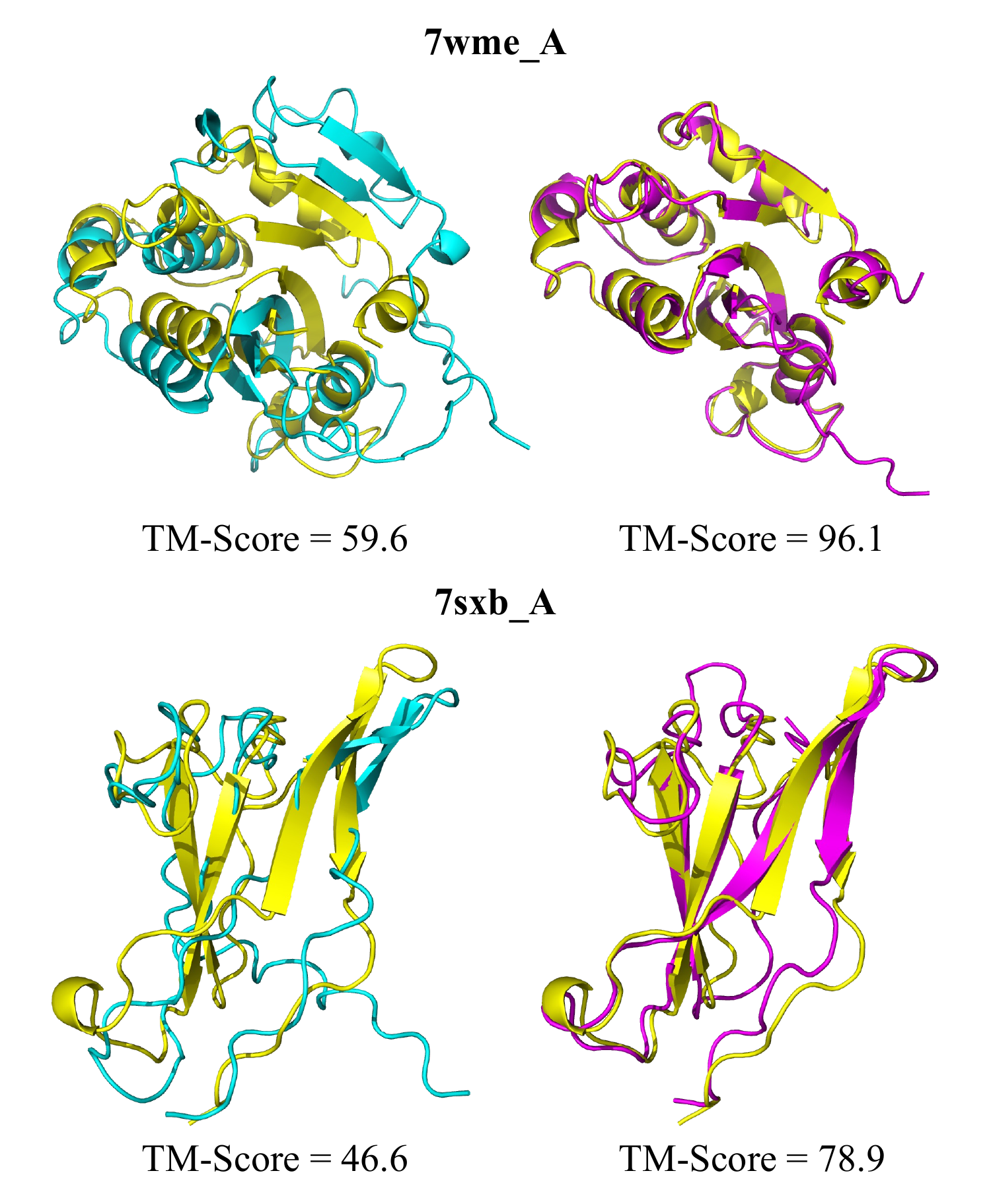}
	% \vspace{-10pt}
	\caption{\label{fig:dpo3b} \textbf{Visualization of improved structure prediction after DPO.}
    \textmd{\textcolor{yellow}{Yellow}: Ground truth; 
    \textcolor{cyan}{Blue}: Predictions based on MSA generated by \model;
    \textcolor{violet}{Pink}: Predictions based on MSA generated by \model-DPO.;}} 
	% \vspace{-10pt}
\end{figure}

\begin{figure*}[t]
	\centering
	\subfigure[generated by \model]{\label{subfig:7wme_3b_msa}
		\includegraphics[width=0.95\textwidth]{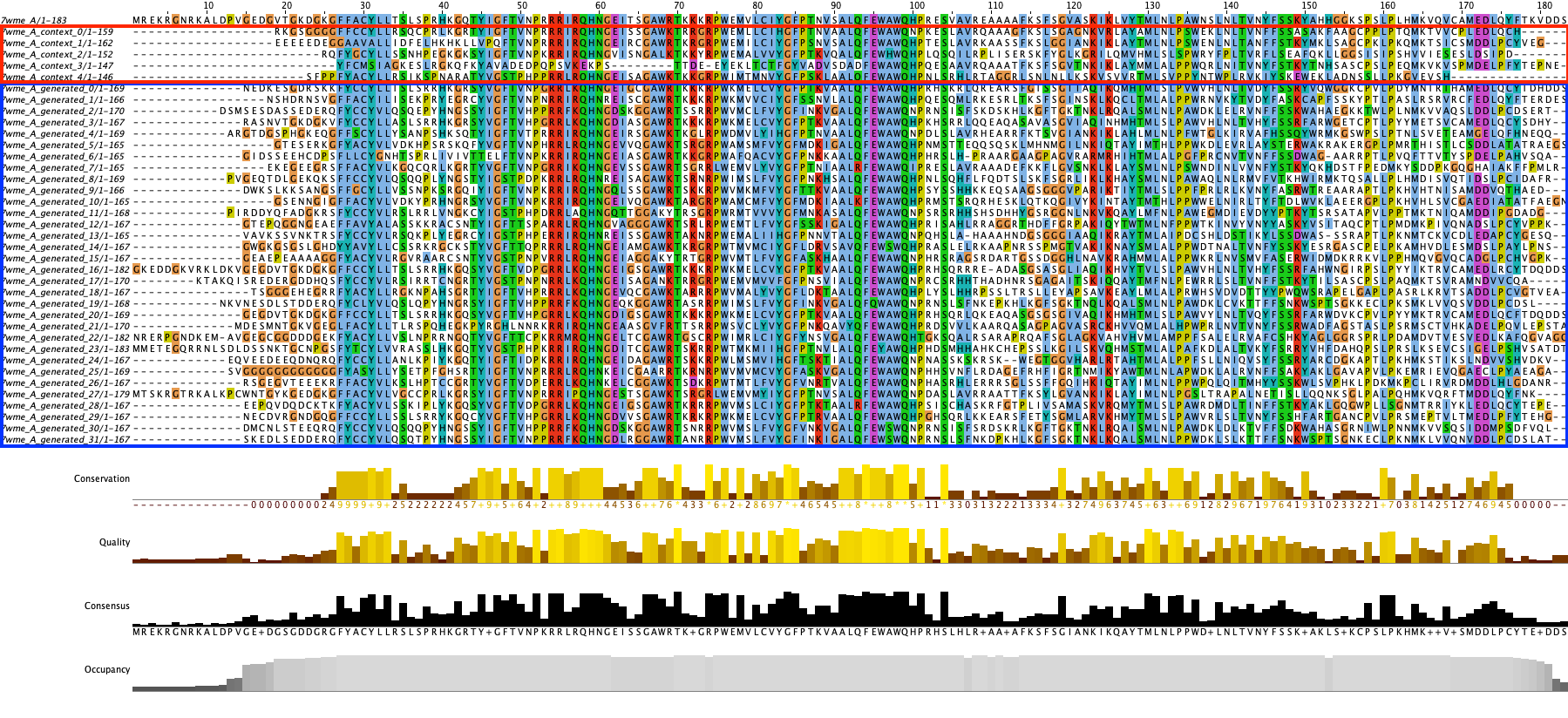}
	}
	% \hspace{-0.1in}
	\subfigure[generated by \model-DPO]{\label{subfig:7wme_dpo_msa}
		\includegraphics[width=0.95\textwidth]{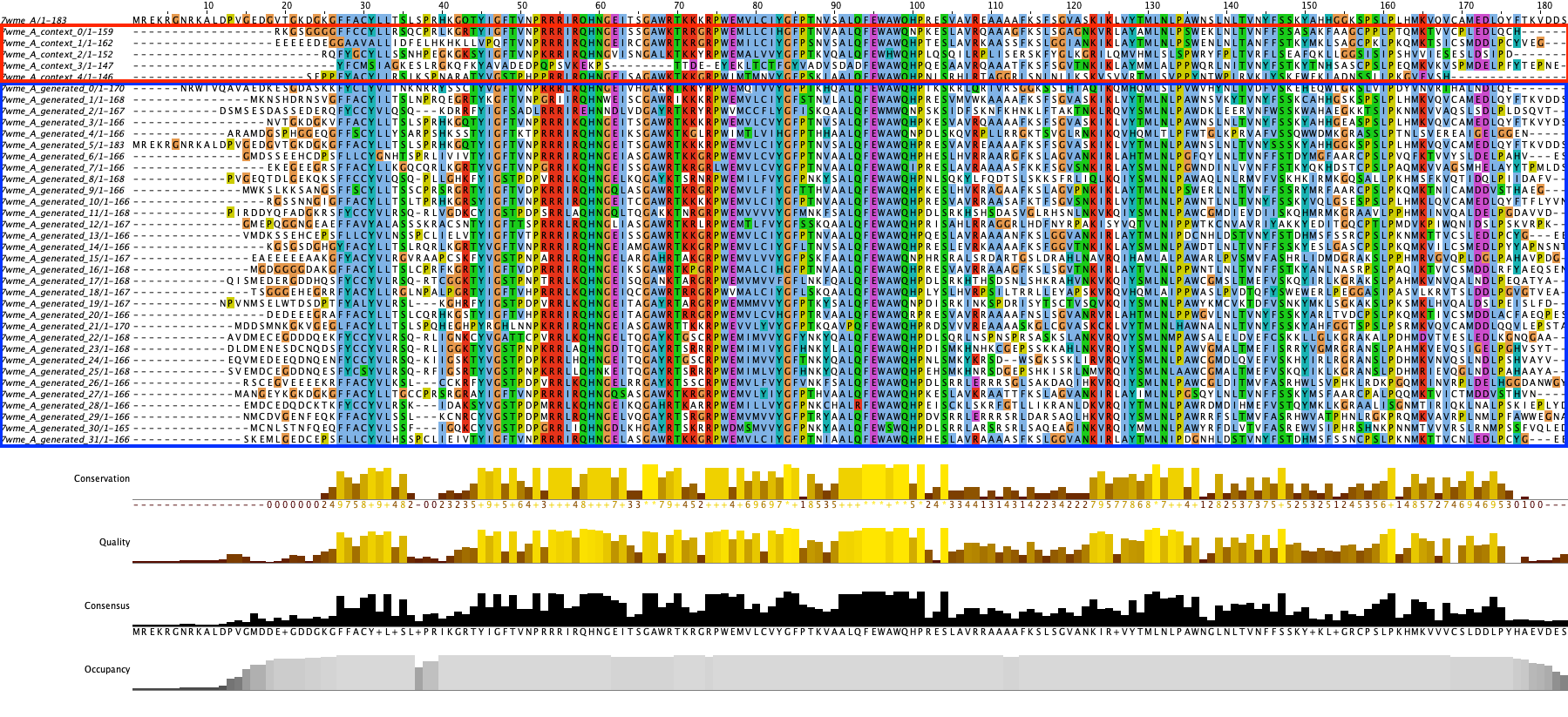}
	}
	% \vspace{-10pt}
	\caption{\label{fig:abla_vs_1} \textbf{Residue Distribution of Generated MSA for 7wme\_A.}
    \textmd{The red box indicates natural MSA used as prompts during generation. The blue box indicates generated MSA. Residues are colored using the clustal scheme by Jalview.}}
	% \vspace{-10pt}
\end{figure*}

\begin{figure*}[t]
	\centering
	\subfigure[generated by \model]{\label{subfig:7sxb_3b_msa}
		\includegraphics[width=0.45\textwidth]{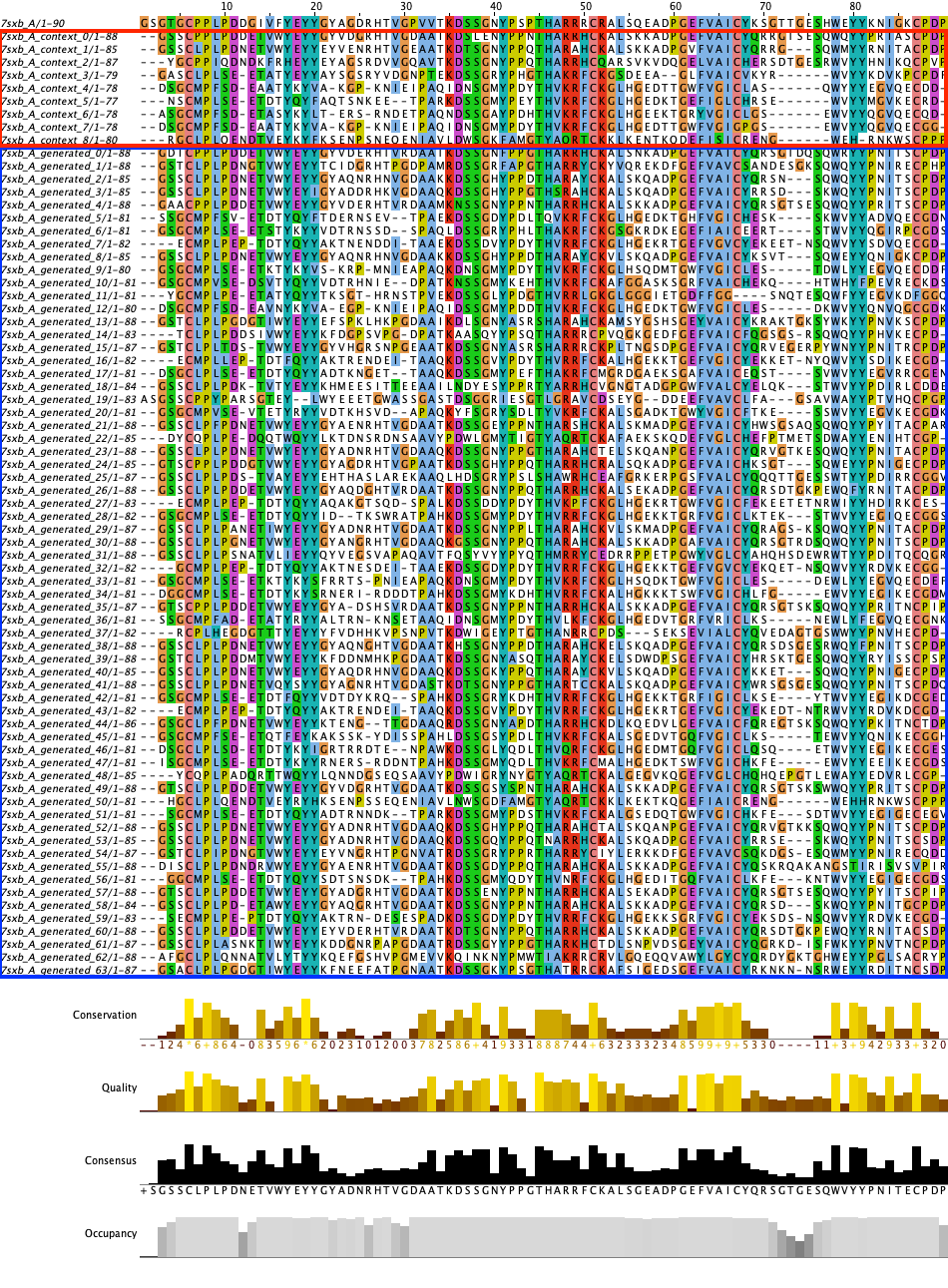}
	}
	\hspace{-0.1in}
	\subfigure[generated by \model-DPO]{\label{subfig:7sxb_dpo_msa}
		\includegraphics[width=0.45\textwidth]{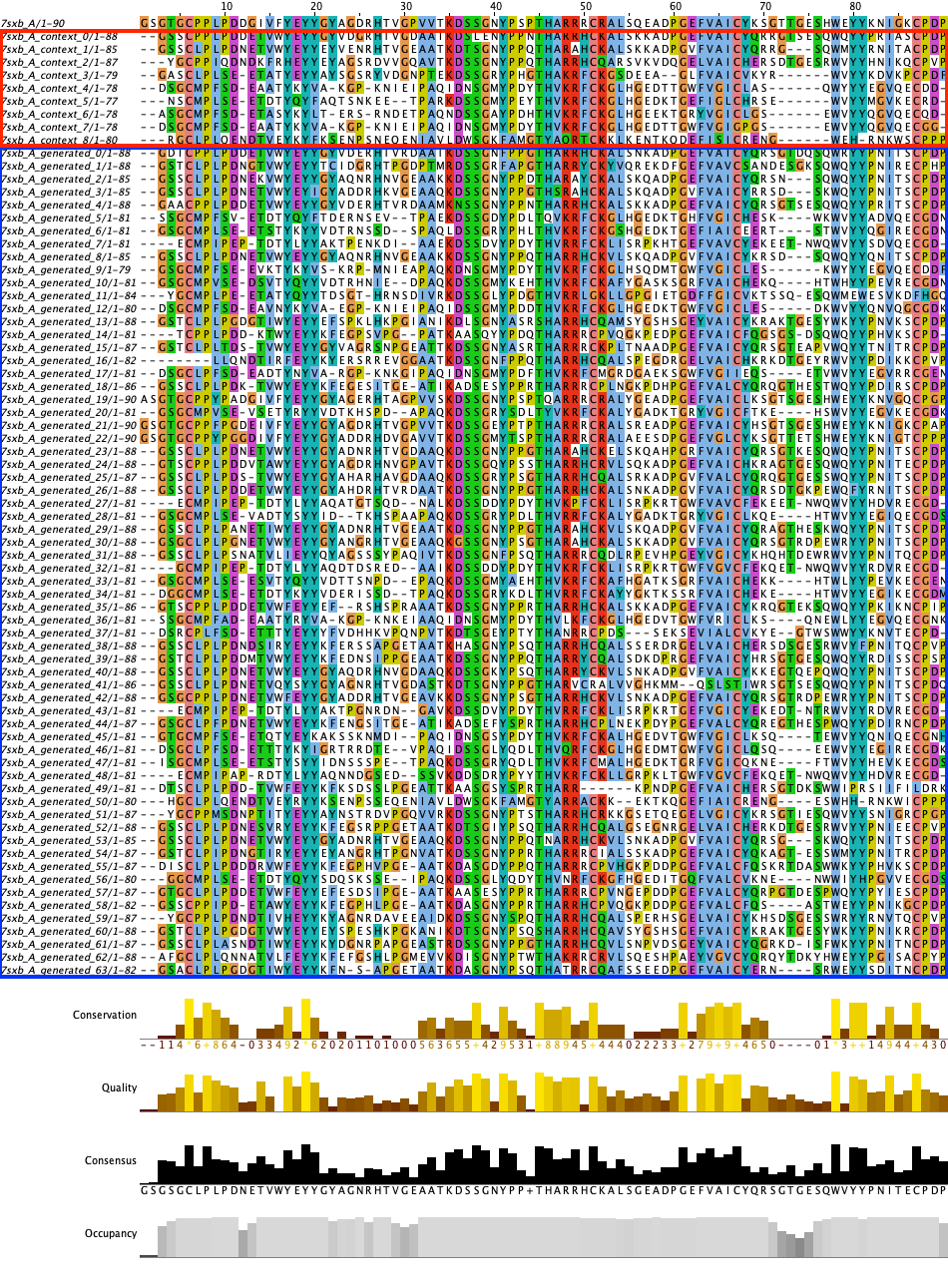}
	}
	% \vspace{-10pt}
	\caption{\label{fig:abla_vs_2} \textbf{Residue Distribution of Generated MSA for 7sxb\_A.}
    \textmd{The red box indicates natural MSA used as prompts during generation. The blue box indicates generated MSA. Residues are colored using the clustal scheme by Jalview.}}
	% \vspace{-10pt}
\end{figure*}

Figure~\ref{fig:dpo3b} represents the comparison before and after the DPO training, depicting notable enhancements in structure prediction accuracy.
Figure~\ref{fig:abla_vs_1} and \ref{fig:abla_vs_2} provide an in-depth analysis of the generated MSA for each case. Specifically, residues 43, 53, 71-79, 105-111, 122, 132 and 157-166 in the MSA of 7wme\_A, along with residues 22-27, 53, and 73 in the MSA of 7sxb\_A, display distinct characteristics pre- and post-DPO training.

% \section{Appendix / supplemental material}

% Optionally include supplemental material (complete proofs, additional experiments and plots) in appendix.
% All such materials \textbf{SHOULD be included in the main submission.}
\end{document}